\documentclass[journal,12pt,one column]{IEEEtran}
\usepackage{amsfonts}
\usepackage{amssymb}
\usepackage{amsmath}
\usepackage{graphicx}
\usepackage{color}
\usepackage{multirow}
\usepackage{cite}
\usepackage{verbatim}
\usepackage{bm}
\usepackage{epstopdf}
\usepackage{subfigure}
\usepackage{pifont}
\usepackage{authblk}
\usepackage{setspace}
\usepackage{geometry}
\usepackage[stable]{footmisc}
\usepackage{color}
\usepackage{stfloats}
\usepackage{algorithm}
\usepackage{algorithmic}
\usepackage{tabularx}
\usepackage{pifont}

\input{tcilatex}
\newtheorem{property}{Property}

\newtheorem{rema}{Remark}

\def\xM0{{\rm M}_0}

\def\xLRm#1{\left[#1\right]}
\def\xLRs#1{\left(#1\right)}
\def\xMset{\Omega}

\def\SNRy{{\rm SNR}_{sim}}
\begin{document}

\title{Doubly-Irregular Repeat-Accumulate Codes over Integer Rings for
Multi-user Communications}
\author[$\#$]{Fangtao Yu, Tao Yang, \emph{Member, IEEE} and Qiuzhuo Chen}
\maketitle

\begin{abstract}
Structured codes based on lattices were shown to provide enlarged capacity
for multi-user communication networks. In this paper, we study
capacity-approaching irregular repeat accumulate (IRA) codes over integer
rings $\mathbb{Z}_{2^{m}}$ for $2^m$-PAM signaling, $m=1,2,\cdots$. Such
codes feature the property that the integer sum of $K$ codewords belongs to
the extended codebook (or lattice) w.r.t. the base code. With it, \emph{%
structured binning} can be utilized and the gains promised in lattice based
network information theory can be materialized in practice. In designing IRA
ring codes, we first analyze the effect of zero-divisors of integer ring on
the iterative belief-propagation (BP) decoding, and show the invalidity of
symmetric Gaussian approximation. Then we propose a doubly IRA (D-IRA) ring
code structure, consisting of \emph{irregular multiplier distribution} and
\emph{irregular node-degree distribution}, that can restore the symmetry and
optimize the BP decoding threshold. For point-to-point AWGN channel with $%
2^m $-PAM inputs, D-IRA ring codes perform as low as 0.29 dB to the capacity
limits, outperforming existing bit-interleaved coded-modulation (BICM) and
IRA modulation codes over GF($2^m$). We then proceed to design D-IRA ring
codes for two important multi-user communication setups, namely
compute-forward (CF) and dirty paper coding (DPC), with $2^m$-PAM signaling.
With it, a physical-layer network coding scheme yields a gap to the CF limit
by 0.24 dB, and a simple linear DPC scheme exhibits a gap to the capacity by
0.91 dB.
\end{abstract}

\affil[$\#$]{School of Electronics Information Engineering, Beihang University, Beijing, 100191, China\authorcr
	Email: \{16711123, tyang, chenqiuzhuo\}@buaa.edu.cn} \renewcommand*{%
\Affilfont} 

\markboth{}{Shell
	\MakeLowercase{\textit{et al.}}: Bare Demo of IEEEtran.cls for IEEE Journals}
\begin{IEEEkeywords}
	Coded modulation, lattice codes, physical-layer network coding, compute-forward, network information theory, multiple-access, broadcast channel, dirty paper coding
\end{IEEEkeywords}

\section{Introduction}

The noisy channel coding theorem reveals the fundamental limits of reliable
communications, and various coding techniques are developed for approaching
the limits. Existing turbo, polar and low-density parity-check (LDPC) codes
can yield near-capacity performance for long block lengths.
Repeat-Accumulate (RA) codes proposed by Divsalar and Jin enjoy both
advantages of linear encoding complexity of turbo codes and parallel
decoding of LDPC codes. Irregular repeat accumulate (IRA) codes feature
non-uniform variable and check nodes degrees which give rise to improved
decoding threshold \cite{jin2000irregular,4215147}. Using density evolution
(DE) or extrinsic information transfer (EXIT) chart based optimization,
well-designed IRA codes perform only a small fraction of dB away from the
capacity limits of binary-input channels \cite%
{khandekar2002irregular,ten2003design}.

For higher order modulation, e.g., $2^{m}$-PAM or $2^{2m}$-QAM, $%
m=1,2,\cdots $, bit-interleaved coded modulation (BICM), trellis-coded
modulation (TCM) and superposition-coded modulation (SCM) have been studied
\cite{caire1998bit,li1999trellis,gadkari1999time}. These conventional
schemes are referred to as \textquotedblleft \emph{binary coding oriented}%
\textquotedblright\ : an off-the-shelf binary channel code is determined in
the first place, and then a \emph{many-to-one} mapping is utilized to match $%
2^{m}$ binary coded digits to a PAM symbol. To approach the capacity limit,
these schemes require an outer-loop receiver iteration\cite{1025496} that
exchanges soft information between the soft-input soft-output demodulator
and a bank of channel-code decoders. As each decoder may involve an
inner-loop iteration by itself, the total number of decoding iterations
amounts to the product of the numbers of inner-loop and out-loop iterations.
Most existing practical systems incline to avoid the outer-loop iteration to
reduce the implementation cost and latency, but at the expense of a
significant gap to the ultimate performance.

Different from the coding-oriented schemes, Chiu proposed $q$-ary IRA
modulation codes for $q$-PAM inputs \cite{chiu2009bandwidth}. This scheme is
referred to as \textquotedblleft \emph{modulation-oriented}%
\textquotedblright : $q$-PAM signaling is determined in the first place, and
an IRA code over GF($q$) is adopted whose output $q$-ary coded digits are
\emph{one-to-one} mapped to $q$-PAM symbols. Thanks to the one-to-one
mapping, the outer-loop iteration is avoided while achieving the
near-capacity performance. Moreover, for prime $q$, IRA modulation codes are
lattice codes without a one-dimension shaping code, whose advance in the
two-way relay channel setup was reported in \cite{yang2015achieving}.



\subsection{Motivations and Necessity of Ring Codes in Multi-user Networks}

For a variety of multi-user configurations, structured codes based on
lattices have been exploited in solving network information theory problems
\cite{li2003linear}, such as Slepian-Wolf and Wyner-Ziv problems (source
coding with side information (SI) at receiver), dirty paper coding (DPC)
problem (channel coding with SI at transmitter) \cite%
{zamir2002nested,erez2005close}, physical-layer network coding (PNC) or
compute-and-forward (CF) \cite{nazer2011compute}, interference alignment,
multiple-access (MA), precoding for broadcast channel, and etc.. Using
lattices codes, compelling theoretical advances by exploiting
\textquotedblleft \emph{structured binning}\textquotedblright\ over
conventional random coding have been reported, where the key notion is to
efficiently compute the \emph{bin-indices} \cite%
{CadambeIT08,NtranosISIT13,YangTWC17}. The proofs of these results were
based on the existence of \textquotedblleft Roger-good\textquotedblright\
and \textquotedblleft Ployrev-good\textquotedblright\ lattice chains \cite%
{erez2005close}, but no clues are given on the code construction for
practical implementation.

To materialize the gains of structured binning in a practical multi-user
wireless network with widely used $q=2^{m}$ level PAM (or $2^{2m}$-QAM)
signaling, codes over integer rings $\mathbb{Z}_{2^{m}}$ become particularly
relevant. To see this, first note that conventional BICM, TCM and SCM
schemes are not lattice codes. Due to the many-to-one signal mapping,
structured binning does not apply therein. Second, the aforementioned IRA
modulation codes belong to lattice codes only for prime $q$. Yet, for
non-prime $q=2^{m}$, the IRA modulation codes operate over the extended
Galois field GF($2^{m}$) \cite{chiu2009bandwidth}. The additive and
multiplication rules of GF$\left( 2^{m}\right) $ are not identical to the
integer operations of $\mathbb{Z}_{2^{m}}$, hence structured binning does
not apply, either.\emph{\ }This motivates us to study ring codes over
integers $\mathbb{Z}_{2^{m}}$.

\subsection{Main Contributions}

To the best of our knowledge, the design of capacity-approaching ring codes
with $2^{m}$-PAM signaling remains open. In this paper, we first analyze the
effect of zero-divisor elements in $\mathbb{Z}_{2^{m}}$ on the
belief-propagation (BP) decoding. We show the invalidity of the symmetric
Gaussian approximation (with which the results in \cite{chiu2009bandwidth}
are built) in the statistics of the soft information exchanged in the
component decoders. Then, we propose a new doubly IRA (D-IRA) ring code,
featuring \emph{irregular multiplier distribution} and \emph{irregular
node-degree distribution}, that can restore the symmetry and optimize the
decoding threshold. The degree profile optimization based on extrinsic
information transfer chart (EXIT) curve-fitting is conducted \cite{1494432}.
We demonstrate that our proposed D-IRA ring codes perform as low as 0.29 dB
away from the AWGN capacity limits with $2^{m}$-PAM inputs, and outperform
other baseline code-modulation schemes.

We then move on to the design of D-IRA ring codes for the CF and DPC
settings operated with structured binning \cite%
{nazer2011compute,ZhuIT17,YangTWC17_NOMA, YangTWC14}, with $2^{m}$-PAM
signaling. With it, it is shown that the D-IRA ring-coded PNC yields a gap
to the CF capacity limit by 0.24 dB, and a simple linear DPC scheme exhibits
a gap to the interference-free capacity by as low as 0.91 dB. D-IRA ring
codes may serve as a bridging between the lattice-based network information
theory and practical wireless systems.

This paper focuses on designing ring codes of $2^m$-PAM signaling that
achieve the near-capacity performance of some multi-user communication
setups, hence the decoding thresholds (waterfall region) with long codes are
primarily concerned. The code profiles optimized for long codes are also
competitive choices for medium-length codes. The design of short codes
require distance spectrum and weight analysis over a $q$-ary ring. This is
out of the scope of the current paper and will be considered as a future
work.

\section{Preliminaries of $2^m$-ary Codes over Integer Rings}

Throughout this paper we present the real-valued model with $2^{m}$-PAM. The
complex-valued model with $2^{2m}$-QAM can be easily represented by a
real-valued model of doubled dimension as treated in \cite{nazer2011compute}
\cite{yang2015achieving}.

\subsection{Ring Codes for $2^m$-PAM Signaling}

Let $\mathbf{w=}\left[ w_{1},\cdots ,w_{k}\right] ^{T}$ denote a $2^{m}$-ary
message sequence of length $k$\footnote{%
The conversion from a binary message sequence to a $2^m$-ary message
sequence is straightforward.}. Each entry of $\mathbf{w}$ belongs to an
integer ring $%
\mathbb{Z}
_{2^{m}}\triangleq \left\{ 0,1,\cdots ,2^{m}-1\right\}$. A $2^{m}$-ary ring
code with generator matrix $\mathbf{G}$ is employed to encode $\mathbf{w}$,
given by%
\begin{equation}
\mathbf{c}=\mathbf{G\otimes w}  \label{Eq_encodinggeneral}
\end{equation}%
where \textquotedblleft $\mathbf{\otimes }$\textquotedblright\ represents
matrix multiplication modulo-$2^{m}$. The generator matrix $\mathbf{G}$ is
of size $n$-by-$k$ with entries in $\mathbf{\in }$ $%
\mathbb{Z}
_{2^{m}}$. Let $\mathcal{C}^{n}$ denote the codebook which collects all
valid codewords of $\mathbf{c}$ generated by (\ref{Eq_encodinggeneral}).

A random vector $\mathbf{\theta}$ $\in $ $%
\mathbb{Z}
_{q}^{n}$ is generated and added on $\mathbf{c}$, resulting in $\mathbf{c}%
^{\prime }=\mathbf{c}\oplus \mathbf{\theta}$ where \textquotedblleft $\oplus
$\textquotedblright\ represents the matrix addition modulo-$2^{m}$. This is
for the purpose of random permutation \cite{chiu2009bandwidth}. Then, each
entry of $\mathbf{c}^{\prime }$ is \textit{one-to-one}\ mapped to a symbol
that belongs to a constellation of $2^{m}$ points. For $2^{m}$-PAM
constellation with uniformly spaced points, the mapping function $\delta
(\cdot )$ is simply
\begin{equation}
\mathbf{x}=\delta \left( \mathbf{c}^{\prime }\right) =\frac{1}{\gamma }%
\left( \mathbf{c}^{\prime }-\frac{2^{m}-1}{2}\right) \in \frac{1}{\gamma }%
\left\{ \frac{1-2^{m}}{2},\cdots ,\frac{2^{m}-1}{2}\right\} ^{n},
\label{Eq_PAMmapping}
\end{equation}%
implemented symbol-wisely. Here $\gamma $ is a normalization factor to
ensure unit average symbol energy. The information rate is $R=\frac{k}{n}%
\log _{2}q=\frac{km}{n}$ bits/symbol.

Roughly speaking, the problem is to find a \textquotedblleft
good\textquotedblright\ structure of $\mathbf{G}$ that achieves
near-capacity, while the encoding, decoding and code optimization can be
implemented with a reasonable cost.


\begin{rema}
The ring coded $2^{m}$-PAM scheme differs from conventional coding-oriented\
schemes, where binary coded sequence $\mathbf{c}$ is de-multiplexed into $m$
streams $\mathbf{c}^{(1)},\cdots ,\mathbf{c}^{(m)}$. Then, a \textit{%
many-to-one} mapping is employed, e.g. the Grey mapping used in BICM. Such a
many-to-one mapping incurs uncertainty that has to be addressed in the first
place at the receiver.
\end{rema}

%

\begin{property}
For any $K$ codewords $\mathbf{c}_{1},\mathbf{c}_{2},\cdots ,\mathbf{c}%
_{K}\in \mathcal{C}^{n}$, the ring coded $2^{m}$-PAM scheme satisfies
\begin{equation}
\func{mod}\left( \sum_{i=1}^{K}\alpha _{i}\mathbf{c}_{i},2^{m}\right) \in
\mathcal{C}^{n}
\end{equation}%
for any integer coefficients $\left[ \alpha _{1},\cdots ,\alpha _{K}\right] $%
. In other words, the integer-sum of $K$ codewords modulo-$2^{m}$ remains as
a valid codeword, hence the name \textquotedblleft \textit{integer} \textit{%
additive property\textquotedblright }.
\end{property}

This property\ has been intensively studied in the area of lattice codes for
solving network information theory problems \cite%
{zamir2002nested,nazer2011compute,21246}. The details will be retained until
Section V. This property does not hold in conventional binary coded-oriented
schemes.

\subsection{Rings Versus Galois Fields}

Most existing works on lattice codes, low density lattice codes, and IRA
modulation codes focused on prime $q$ \cite{SommerTIT08}\cite%
{chiu2009bandwidth}, where GF$\left( q\right) $ and $%
\mathbb{Z}
_{q}$ are equivalent. The integer additive property holds therein. In
practical systems utilizing BPSK to 4096-QAM signaling, non-prime $q$ $=2^{m}
$ is required. The operation rules of $\mathbb{Z}_{2^{m}}$ are different to
those of GF$\left( 2^{m}\right) $, and integer additive property does not
hold for GF$\left( 2^{m}\right) $ based codes. To see this, recall that GF$%
\left( 2^{m}\right) $ is an extension field of GF$\left( 2\right) $, which
has elements $\left\{ 0,1,\beta ,\beta ^{2},\cdots \beta ^{2^{m}-2}\right\} $
\cite{LinShuTextbook}. The additive rule w.r.t. these elements is determined
based on the primitive element of the polynomials, which is different from
that of $\mathbb{Z}_{2^{m}}$. Therefore, to enable the integer additive
property for $2^{m}$-PAM signaling, utilization of ring codes over $%
\mathbb{Z}
_{2^{m}}$ could be a must.


The ring coded $2^{m}$-PAM is a simplified yet powerful version of nested
lattice codes whilst the GF$\left( 2^{m}\right) $ based codes are not. The
fine lattice is given by the \emph{extended codebook} w.r.t. $\mathbf{c}=%
\mathbf{G\otimes w}$. This is also referred to as \textquotedblleft
Construction A\textquotedblright\ of lattice codes \cite{nazer2011compute},
\cite{21246}. The shaping lattice is given by $2^m\mathbb{Z}^{n}$, i.e., a
one-dimension modulo-$2^{m}$ operation, which yields $2^m$-PAM signaling.
Note that this paper devotes no efforts to attain a Gaussian input
distribution, although this can be interesting additive future works.
Comparing to Gaussian signaling, $2^m$-PAM enjoys lower implementation cost
and lower peak-to-average power ratio (PAPR) that favours practical
implementation.

\section{Proposed Doubly-Irregular Repeat Accumulate Ring Codes}

\subsection{Zero-divisors in Integer Rings}

Recall the integer ring $\mathbb{Z}_{2^{m}}=\left\{ 0,1,\cdots
,2^{m}-1\right\} $ where the addition and multiplication are defined as
\begin{equation}
\begin{aligned} a\oplus b&\triangleq(a+b)\ {\rm mod}\ 2^m,\\ a\otimes
b&\triangleq(a\cdot b)\ {\rm mod}\ 2^m. \end{aligned}  \label{E1}
\end{equation}%
For a non-zero element $a\in \mathbb{Z}_{2^{m}}$, its inverse is said to
exist if there is a unique element $b\in \mathbb{Z}_{2^{m}}$ that satisfies $%
a\otimes b=1$. This unique inverse is written as $a^{-1}$. Not all but some
of the non-zero elements have unique inverses. For a non-zero element $a\in
\mathbb{Z}_{2^{m}}$, its \textit{zero-multiplier} is defined as
\begin{equation}
M_{0}(a)\triangleq \min\limits_{j>0,a\otimes j=0}{j}.  \label{E2}
\end{equation}%
For the elements with unique inverses, $M_{0}(a)=q$. Such elements are
called \textit{regular elements}. For the elements that do not have unique
inverses, $M_{0}(a)<q$. Such elements are called \textit{zero-divisors}. An
example of $\mathbb{Z}_{8}$ is shown in TABLE \ref{table 1}, where $\left\{
1,3,5,7\right\} $ are regular elements while $\left\{ 2,4,6\right\} $ are
zero-divisors.

\begin{table}[h]
\caption{Example of zero-divisors with $q$=8.}
\label{table 1}\centering\renewcommand{\multirowsetup}{\centering} %
\renewcommand{\arraystretch}{1.2} {\small
\begin{tabular}{c|ccccccc}
\hline
$a$ & 1 & 2 & 3 & 4 & 5 & 6 & 7 \\ \hline
$M_{0}(a)$ & 8 & 4 & 8 & 2 & 8 & 4 & 8 \\ \hline
\end{tabular}%
}
\end{table}

\subsection{Repeat Accumulate (RA) Ring Code Structure}

\begin{figure}[h]
\centering\includegraphics[scale=0.35]{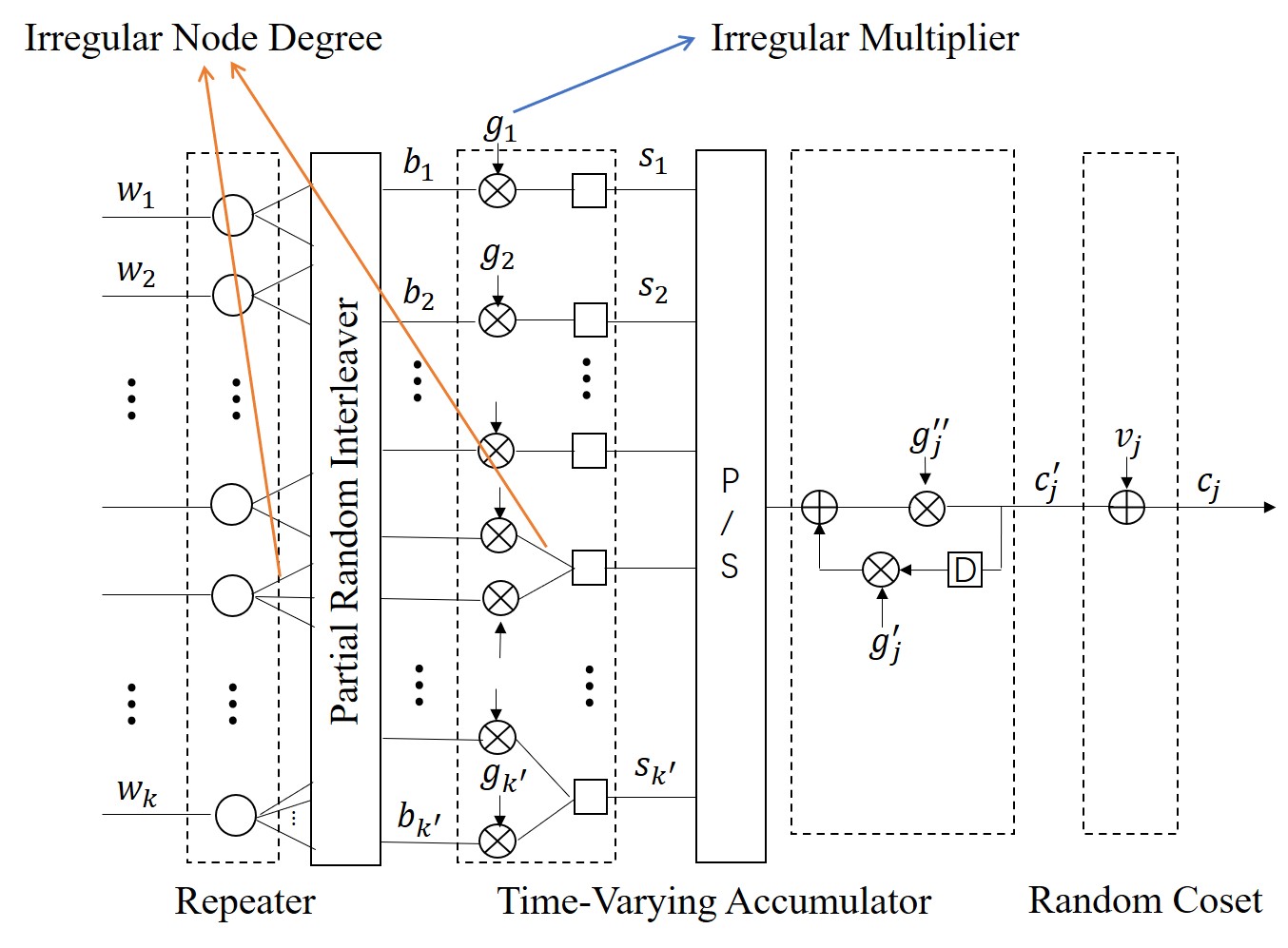}
\caption{Block diagram of a D-IRA code encoder. The multipliers $\mathbf{g}%
^{\prime }$ and $\mathbf{g}^{\prime \prime }$ of the time-varying
accumulator are randomly selected from the regular element set with equal
probability. }
\label{fig8}
\end{figure}
The block diagram of the encoder is depicted in Fig. \ref{fig8}. Each entry
of the message sequence $\mathbf{w}$ is referred to as an information node.
The entries are repeated according to a certain degree distribution, which
yields length-$k^{\prime }$ sequence $\mathbf{b}^{\prime }\in \left\{
0,\cdots ,2^{m}-1\right\} ^{k^{\prime }},k^{\prime }>k$. This sequence is
interleaved, yielding $\mathbf{b}=\pi (\mathbf{b}^{\prime })$. The
interleaved sequence is forwarded to a bank of check-nodes (CNs). Each input
edge of the CNs is associated with a multiplier with value taken in $\left\{
1,\cdots ,2^{m}-1\right\}$. Note that the multipliers can be either regular
elements or zero-divisors. The outputs of the CNs are forwarded to a
time-varying accumulator, which yields the length-$n$ codeword sequence $%
\mathbf{c}$ \cite{chiu2009bandwidth}. Next $\mathbf{c}$ is one-to-one mapped
to $\mathbf{x}$ as given in (\ref{Eq_PAMmapping}). For the clarity of
presentation, we omit the random permutation $\mathbf{\theta}$ .

For AWGN channel, upon receiving $\mathbf{y}$, the noisy observation of $%
\mathbf{x}$, the receiver first calculates the channel-intrinsic symbol-wise
a posteriori probabilities (APPs). Let $c[t]$, $y[t]$, $t=1,\cdots ,n$,
denote the $t$-th entry of $\mathbf{c}$ and $\mathbf{y}$, respectively. The
channel-intrinsic APP for $c[t]=i$, $i=0,\cdots ,2^m-1$, is
\begin{equation}
p_{i}^{CH}\left[ t\right] \triangleq p\left( c[t]=i|y[t]\right) \propto
p\left( y[t]|c[t]=i\right) =\frac{1}{\eta }\exp \left( -\frac{\left(
y[t]-\delta \left( i\right) \right) ^{2}}{2\sigma _{z}^{2}}\right)
\label{Eq_symbolwiseAPP}
\end{equation}%
where the conversion from APP to likelihood function follows from the Bayes
rule as $c[t]$ is uniformly distributed, and $\eta $ is just a normalization
factor to ensure $\dsum\limits_{i}p_{i}^{CH}\left[ t\right] =1$. The $2^m$%
-level probabilities are collected by a probability vector $\mathbf{p}^{CH}%
\left[ t\right] =\left[ p_{0}^{CH}\left[ t\right] ,\cdots ,p_{2^{m}-1}^{CH}%
\left[ t\right] \right] ^{T}$.

\begin{figure}[h]
\centering\includegraphics[scale=0.35]{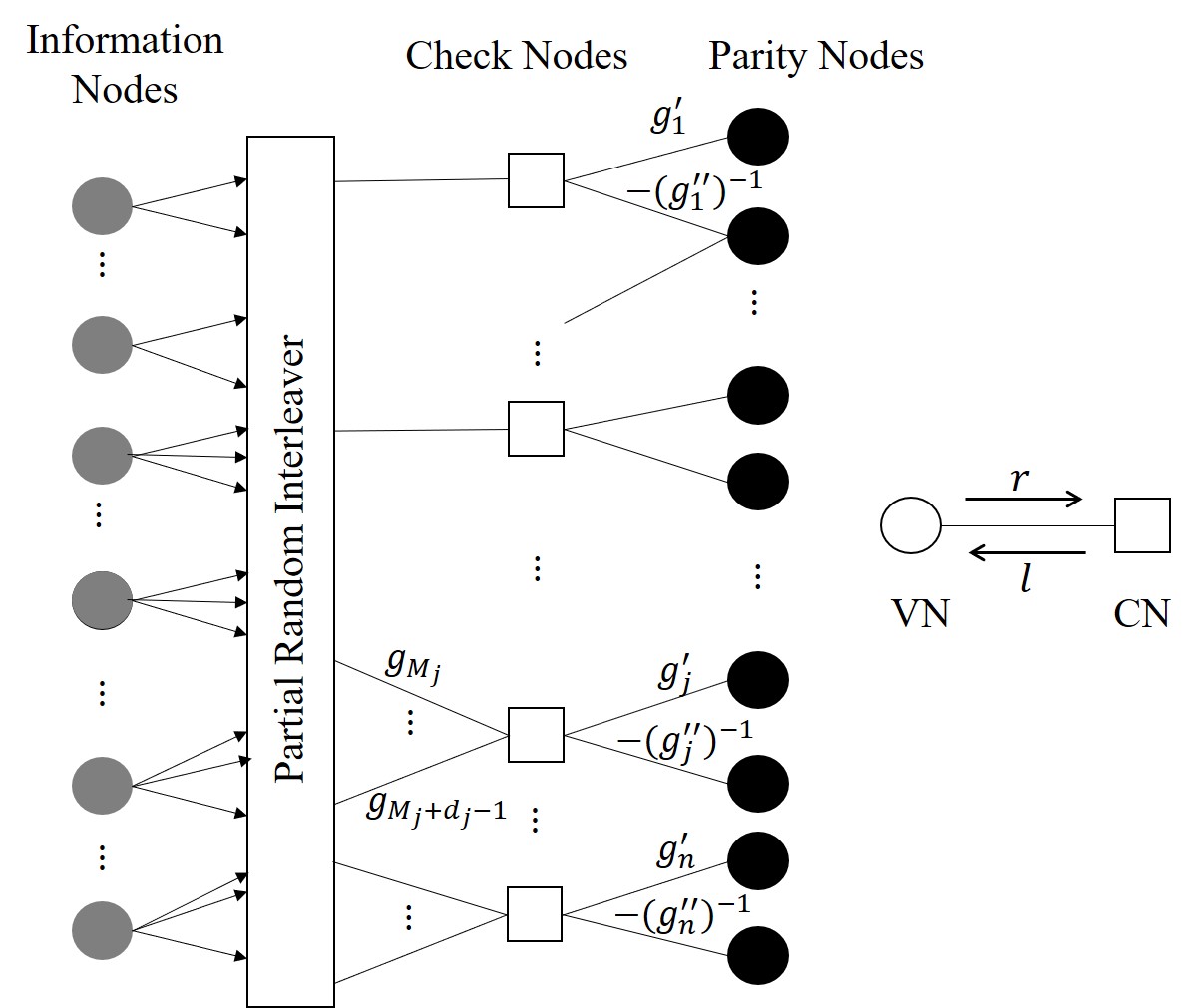}
\caption{Tanner graph of D-IRA.}
\label{Fig_Tanner}
\end{figure}

The APP vectors $\mathbf{p}^{CH}\left[ t\right] ,t=1,\cdots ,n$, are
forwarded to the iterative belief propagation (BP) decoder, which is to
yield the decision on the message sequence $\mathbf{w}$. The Tanner graph of
a D-IRA ring code is shown in Fig. \ref{Fig_Tanner}. There are two types of
message-propagation in BP algorithm: the messages propagated from variable
nodes (VNs) to check nodes (CNs) and the messages propagated from CNs to
VNs. In a generic RA structure, VNs involve 1) information nodes at the
left-hand side of the interleaver and 2) parity nodes at the right-hand side
of the interleaver which are attached to the channel intrinsic APPs. All
nodes operate with the length-($2^{m}-1$) probability vectors.

The output probability vector of a VN of degree $d_{v}$ is computed as
\begin{equation}
r_{i}=\frac{p_{i}\prod_{\tau =1}^{d_{v}-1}l_{i}^{\tau }}{%
\sum_{j=0}^{2^{m}-1}\left( p_{j}\prod_{\tau =1}^{d_{v}-1}l_{j}^{\tau
}\right) },\quad i=0,\cdots ,2^{m}-1  \label{E4}
\end{equation}%
where $l_{i}^{\tau }$ denotes the probability of $i$ obtained from the $\tau
$-th input edge to VN. Here, $p_{i}=\frac{1}{2^{m}}$ for information nodes
and $p_{i}=p_{i}^{CH}\left[ t\right] $ for parity nodes. Here we omitted the
node index in the presentation.

The outputs of a CN of degree $d_{c}$ are computed as
\begin{equation}
l_{i}=\sum_{a_{1},\cdots ,a_{d-1}\in \Phi }{\prod_{\tau =1}^{d-1}r_{a_{\tau
}}^{\tau }}  \label{E3}
\end{equation}%
with
\begin{equation}
\Phi :a_{1},\cdots ,a_{d-1}\in \mathbb{Z}_{2^{m}},\dbigoplus\limits_{\tau
=1}^{d-1}h_{\tau }a_{\tau }\oplus h_{0}i=0,  \label{Eq_CheckRule}
\end{equation}%
denoting the check-rule. Here $d=d_{c}+2$ where the extra two edges are from
the accumulator as in Fig. \ref{Fig_Tanner}; $\left\{ a_{1},\cdots
,a_{d-1}\right\} $ denotes a \textit{candidate symbol-combination} in $%
\left\{ 0,\cdots ,q-1\right\} ^{d-1}$, referred to as a candidate,
satisfying the check-rule constraint; $r_{a_{\tau }}^{\tau }$ denotes the
probability of $a_{n}$ obtained from the $\tau $-th input edge, $h_{\tau }$
is the multiplier of the $\tau $-th edge and $h_{0}$ is that of the output
edge.

In each iteration, the above calculations are carried out and the messages
are exchanged among the VNs and CNs. The iteration stops when the hard
decision of the output sequence meets the constraints of code parity-check
matrix or the maximum number of iterations is reached.

\subsection{Issues with Regular Multiplier Distribution}

\begin{definition}
If the multipliers w.r.t. the $k^{\prime }$ edges, denoted by $\mathbf{g}=%
\left[ g_{1},\cdots ,g_{k^{\prime }\text{ }}\right] ^{T}$, are uniformly
distributed in $\left\{ 1,\cdots ,2^{m}-1\right\} $, i.e.
\begin{equation}
p\left( g=j\right) =\frac{1}{2^m-1},j=1,\cdots,2^m-1,  \label{Eq_RMP}
\end{equation}%
we say that the code has a \emph{regular multiplier distribution}. On the
other hand, if the multipliers are not uniformly distributed, i.e. (\ref%
{Eq_RMP}) does not hold, the code is said to have an \textit{irregular multiplier
distribution}.
\end{definition}

As seen from the code structure, each edge of CNs is associated with a
multiplier $g$ taking values $\left\{ 1,\cdots ,q-1\right\} $. For the $%
2^{m} $-ary ring codes under consideration, there are zero-divisors who have
no unique inverse in $\left\{ 1,\cdots ,q-1\right\} $, causing ambiguity in
the message passing. We next illustrate the impact caused by this ambiguity
issue, and presents how the proposed irregular multiplier distribution
addresses it.

\subsubsection{Effect of zero-divisors on the Calculation at the Check Nodes}

Consider a CN of degree $d_{c}$. There are totally $d=d_{c}+1$ input edges
and one output edge in the calculation. Recall that the multipliers w.r.t
the input edges are given by $[h_{1},\cdots ,h_{d}]$ and that w.r.t. the
output edge is given by $h_{0}$, respectively. Let us temporarily consider
that only the $\tau ^{\prime }$-th input edge has a multiplier $h_{\tau
^{\prime }}$ which is a zero-divisor, while the multiplier of the output
edge $h_{0}$ is a regular element. For given $\left[ a_{1},\cdots ,a_{\tau
^{\prime }-1},a_{\tau ^{\prime }+1},\cdots ,a_{d}\right] $ and $i$, there
are multiple values of $a_{\tau ^{\prime }}$ that satisfy the check-rule
\begin{equation}
h_{\tau ^{\prime }}a_{\tau ^{\prime }}\oplus \dbigoplus\limits_{\tau =1,\tau
\neq \tau ^{\prime }}^{d-1}h_{\tau }a_{\tau }\oplus h_{0}i=0.
\label{Eq_CheckRule2}
\end{equation}%
This ambiguity leads to a larger number of valid candidates which are
involved in the calculation of (\ref{E3}). If there are more than one input
edges whose multipliers are zero-divisors, the impact of the ambiguity
becomes more significant, and a even larger number of valid candidates are
involved in the calculation of (\ref{E3}). This results in an effect of
loosen constraint of CNs.

Next, consider that only the output edge has a multiplier $h_{0}$ of a
zero-divisor, while all input edges have multipliers with regular elements.
For given $\left[ a_{1},\cdots ,a_{d}\right] $, there are multiple values of
$i$ that satisfy the check-rule (\ref{Eq_CheckRule2}). The output
probabilities w.r.t. these values have to be set to be identical. This
increases the uncertainty in the output probability vector and reduces
mutual information. If both an input edge and the output edge have
zero-divisor multipliers, the ambiguity becomes even more difficult to track.

\subsubsection{Asymmetry in the LLR Vector}

For any edge, let the associated probabilities be denoted by a vector $%
\mathbf{p=}\left[ p_{0},\cdots ,p_{2^{m}-1}\right] ^{T}$. The log-likelihood
ratios (LLRs) associated with the elements in $\mathbf{p}$ are defined as
\begin{equation}
\lambda _{i}=\log (p_{0}/p_{i}),i=1,\cdots ,2^{m}-1.
\end{equation}%
We refer to $\mathbf{\lambda =}\left[ \lambda _{1},\cdots ,\lambda _{2^{m}-1}%
\right] ^{T}$ as a LLR vector.

Let $\mathbf{\Lambda }$ collect the LLR vectors of all edges. The input LLR
matrix to the check node decoder (CND), denoted by $\mathbf{\Lambda }%
_{CND,in}$, is assumed to be jointly consistent Gaussian distributed \cite%
{ten2001convergence,Bennatan2006Design} with parameter $\sigma ^{2}$. (This
assumption is supported via extensive simulations.) That is, their mean and
cross-covariance are $\sigma ^{2}/2$ and auto-correlation are $\sigma ^{2}$.
The output LLR vector is denoted by $\mathbf{\Lambda }_{CND,out}$ obtained
via (\ref{E3}).

\begin{figure}[h]
\centering
\includegraphics[scale=0.125]{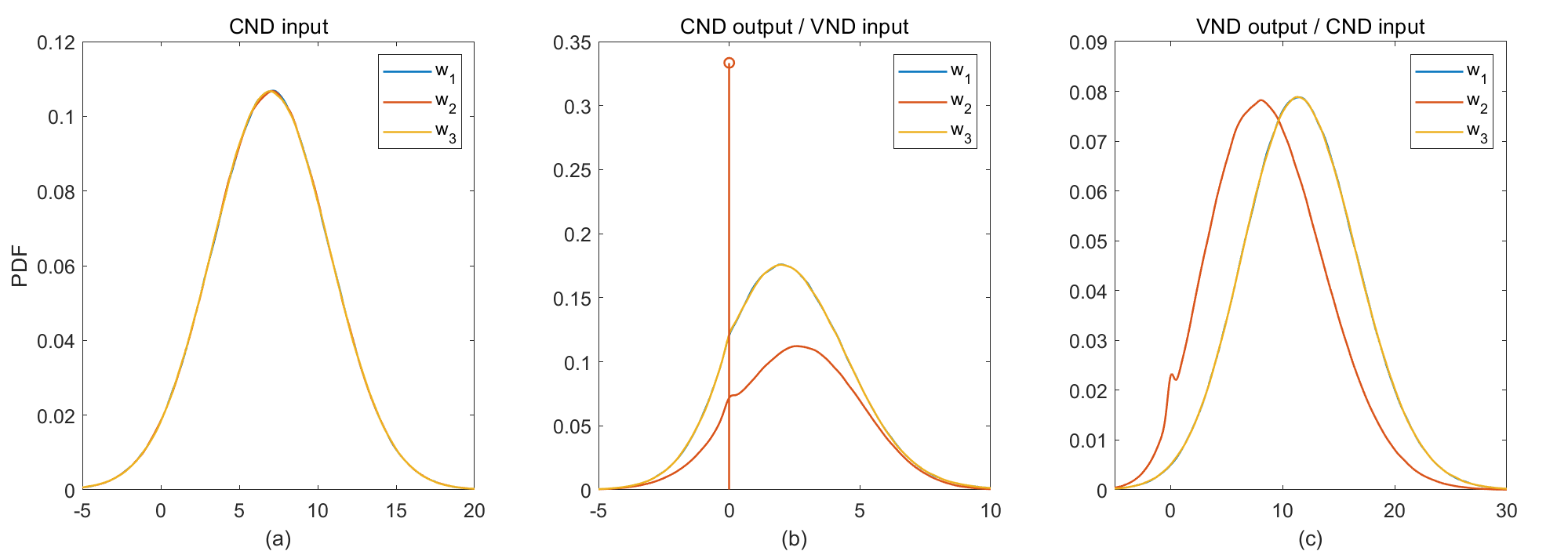}
\caption{PDF of LLR vectors for $m=2$ and multipliers $\mathbf{g}$ uniformly
distributed in $\mathbb{Z}_{4}\backslash \{0\}$.}
\label{fig2}
\end{figure}

For GF($q$) codes with prime $q$, the output LLR vector is symmetric in the
statistics. All the LLR elements follow the same Gaussian p.d.f., at both
the VND and CND. For $2^{m}$-ary ring codes, things start to change.
Consider $m=2$. For RA ring codes with regular multiplier distribution,
i.e., the multipliers in $\mathbf{g}$ are i.i.d. over $\mathbb{Z}%
_{2^{m}}\backslash \{0\}$, the p.d.f. for the LLR vectors are shown in Fig. %
\ref{fig2}. For the output edges with multipliers of value $2$, which is a
zero-divisor, the CND output probability is subject to $p_{0}=p_{2}$, which
results in a zero LLR value $\lambda _{2}=0$. Thus, the p.d.f. corresponding
to $\lambda _{2}$ at the CND output has an impulse at value of zero. The
height of the impulse is given by the ratio between the number of edges with
multiplier 2 and the number of all edges. For the case with regular
multiplier distribution, it is equal to $\frac{1}{3}$. This results in a
different mean of $\lambda _{2}$ relative to that of $\left\{ \lambda
_{1},\lambda _{3}\right\} $.

The CND's output with such p.d.f. is forwarded to the VND. The output of VND
exhibits \textit{asymmetry} between the LLRs for \{$w_{1},w_{3}$\} and $%
w_{2} $. Due to such asymmetric behavior, it is not possible to characterize
the EXIT function via a single-dimension representation, and hence it
requires to utilize the 2-D EXIT chart curve fitting method to optimize the
code degree profiles.

\subsection{Proposed D-IRA Can Restore the Symmetry in Soft information}

\begin{definition}
A repeat-accumulate ring code with both
irregular multiplier distribution and\textit{\ }irregular node degree
distributions is said to have a \textit{doubly-irregular repeat accumulate}
(D-IRA) structure.
\end{definition}

\begin{figure}[h]
\centering
\includegraphics[scale=0.125]{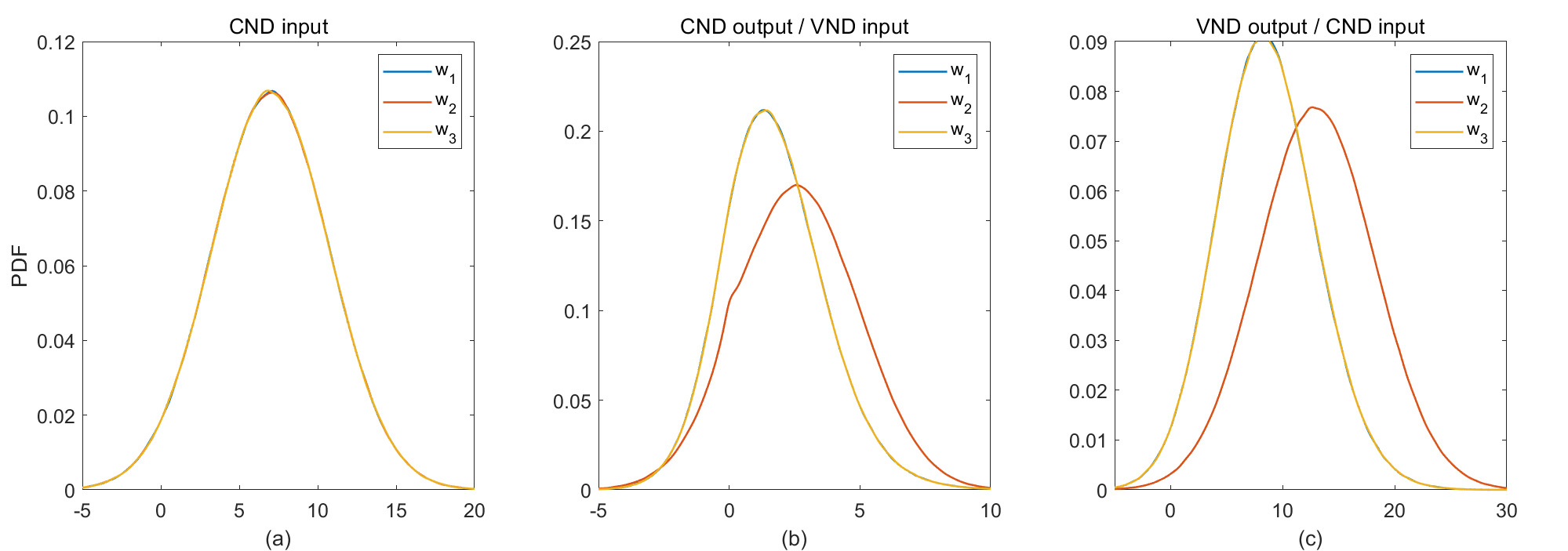}
\caption{PDF of LLR vectors $\mathbf{g}$ uniformly distributed over regular
elements.}
\label{fig3}
\end{figure}

So far we witnessed the impact of the zero-divisors on the asymmetry of the
LLRs. One may expect that using only regular elements in the multipliers can
address this issue, but this is not true. It is shown in Fig. \ref{fig3}
that avoiding using zero-divisors as multipliers will cause a even larger
mean of $\lambda _{2}$.

The idea is to exploit D-IRA structure, that is, to find the irregular
multipliers distribution so that the asymmetry in the LLRs can be restored.
For the $m=2$ example, note that the fraction of the zero-divisor determines
the impulse height of the p.d.f. of $\lambda _{2}$ at value zero. This leads
to a reduced mean of $\lambda _{2}$. By reducing the portion of zero-divisor
2, the mean of $\lambda _{2}$ can be made identical to that of $\lambda _{1}$
and $\lambda _{3}$. It is found that the symmetric Gaussian approximation
can be (approximately) restored in this manner.

For example, consider the irregular multiplier distribution of
[0.4002,0.1996,0.4002] for $g=\left\{ 1,2,3\right\} $, respectively, where
the CND has a degree $d_{c}=3$. The p.d.f. of the LLRs are shown in Fig. \ref%
{fig4}. Then, the output of CND is of identical means for $\left\{ \lambda
_{1},\lambda _{2},\lambda _{3}\right\} $. The VND output vector has (almost)
identical means, even if the p.d.f. of $\lambda _{2}$ is not Gaussian
distributed. Moreover, numerical results show that the variance of VND
output LLRs also have almost identical consistent Gaussian distributions for
$\left\{ \lambda _{1},\lambda _{2},\lambda _{3}\right\} $. The symmetry is
maintained after a complete iteration with this choice of irregular
multiplier distribution. Note that as CN degree varies, the irregular
multiplier distribution that restores the symmetry also varies. In the next
section, this idea will be exploited to find the optimized degree profile of
D-IRA ring codes.

\begin{figure}[h]
\centering
\includegraphics[scale=0.125]{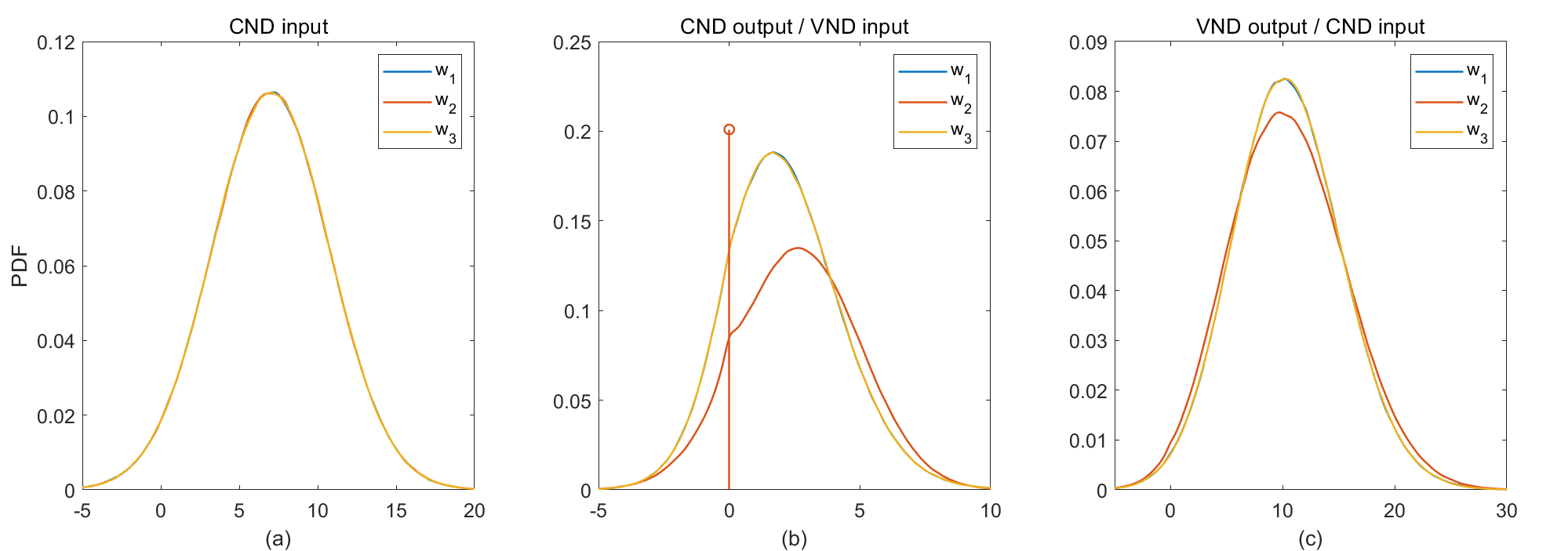}
\caption{PDF of LLR vectors $\mathbf{g}$ distributed over zero-divisors and
regular elements with a specific distribution. The output of CND has
identical mean, and the output of VND has almost identical consistent
Gaussian distribution.}
\label{fig4}
\end{figure}

\section{Optimized Design of D-IRA Ring Codes for $2^{m}$-PAM}

In this section, we optimize the proposed D-IRA ring codes, aiming at
approaching the capacity limit for any $2^m$-PAM signaling. At the current
stage, joint optimization of the irregular multiplier distribution and
irregular node degree distribution is a prohibitive task. We take a
pragmatic approach where the optimization of these two types of
distributions are decoupled.

\subsection{Optimization of Irregular Multiplier Distribution}

The non-zero elements of the integer ring, denoted by $\mathbb{Z}%
_{2^{m}}\backslash \{0\}$, is partitioned as follows. Let a subset $\Omega
_{j}$ collect the Type-$j$ elements in $\mathbb{Z}_{2^{m}}\backslash \{0\}$,
given by
\begin{eqnarray}
\Omega _{0} &\triangleq &\left\{ a\in \mathbb{Z}_{2^{m}}\backslash
\{0\}:M_{0}(a)=2^{m}\right\} ,  \notag \\
\Omega _{j} &\triangleq &\left\{ a\in \mathbb{Z}_{2^{m}}\backslash
\{0\}:M_{0}(a)=m_{j}\right\} ,j=1,\cdots \,,T-1
\end{eqnarray}%
where $T$ denotes the number of types of zero-divisors. The index $j$ is
sorted according to the descending order of $m_{1},\cdots ,m_{T}$. The
cardinality $|\Omega _{j}|$, represents the number of different
zero-divisors of type $j$.

\example Recall the zero-divisors depicted in Table \ref{table 1} with $%
2^{m}=8$. The elements in $\mathbb{Z}_{8}$ are grouped into three types. The
elements \{2,6\} have the same zero-multiplier $M_{0}=4$, referred to as
Type-I zero-divisors (or Type-I elements), where $|\Omega _{1}|=2$.\ The
element \{4\} has zero-multiplier $M_{0}=2$, referred to as Type-II
zero-divisor (or Type-II elements), where $|\Omega _{2}|=1$. The elements
\{1,3,5,7\} are the regular elements whose zero-multiplier is $M_{0}=2^{m}$,
which is referred to as Type-0 elements.

%


Let $N_{d_{c}}$ be the total number of degree-$d_{c}$ CNs. Let the
multiplier distribution w.r.t. different types of zero-divisors be denoted
by $\widetilde{\mathbf{p}}=\left[ \widetilde{p}_{0},\cdots \,,\widetilde{p}%
_{T-1}\right] ^{T}$, where $\widetilde{p}_{j}$ is the probability of Type-$j$
multipliers taking values in $\Omega _{j}$, $\dsum\limits_{j=0}^{T-1}%
\widetilde{p}_{j}=1$. Due to the symmetry among the zero-divisors of the
same type, the individual zero-divisors in $\Omega _{j}$ are allocated with
equal probability. For a multiplier $a\in \Omega _{j}$, its probability is $%
\frac{\widetilde{p}_{j}}{|\Omega _{j}|}$.

The basic notion of the optimization is to equalize the means of the LLRs at
the CND's output for all $T$ types of elements. With it, the symmetry of the
LLRs is preserved at the VND's output shown in numerical results, which
validates the EXIT curve-fitting. Let $\mathbf{\Lambda }_{in}$ be a $%
N_{d_{c}}d_{c}$ by ($2^{m}-1$) dimension LLR matrix as input to the CNs and $%
\mathbf{\Lambda }_{out}$ be that of CN output. Consider the all-zero
codeword $\mathbf{c}=\mathbf{0}$ and a random coset $\mathbf{\theta }$. The
corresponding multiplier sequence $\mathbf{g}=[g_{1},\cdots
,g_{N_{d_{c}}d_{c}}]$ is generated according to a given distribution $%
\widetilde{\mathbf{p}}$. Using (\ref{E4}), (\ref{E3}), the CNs' output LLR
matrix $\mathbf{\Lambda }_{out}$ is obtained.

Let $\lambda _{i,j}$ be the ($i,j$)-th entry of matrix $\mathbf{\Lambda }%
_{out}$, it can be divided into partitions with $\xM0(g_{i})$ and $\xM0(j)$.
The mean matrix $\mathbf{\Theta }=[\vartheta _{r,s}]_{T\times T}$ is
calculated with
\begin{equation}
\vartheta _{r,s}=\left\{ \overline{\lambda }_{i,j}\mid \xM0(g_{i})=m_{r},\xM%
0(j)=m_{s}\right\} .
\end{equation}

The goal is to find a multiplier distribution so that the mean of LLR random
vector $\mathbf{\Lambda }_{out}$ is equal for all types of elements. This is
equivalent to finding%
\begin{equation}
\widetilde{\mathbf{p}}:\widetilde{\mathbf{p}}^{T}\mathbf{\Theta }=\gamma
\mathbf{1}_{1\times T}  \label{Eq_FormulationOpt}
\end{equation}%
We next present an iterative algorithm to find the solution to (\ref%
{Eq_FormulationOpt}) as shown in Algorithm 1 below.

\begin{algorithm}
	\caption{Solving the optimal multiplier distribution}
	\label{alg1}
{\small
	\begin{algorithmic}
		\REQUIRE $d_c$, $I_{A,CND}$, $\SNRy$ and large enough $N_c$
		\ENSURE $\mathbf{p}_{m}^{T}\mathbf{A}=\alpha\mathbf{1}_{1\times (T+1)}$
		\STATE $p_j \gets |\xMset_j|/(q-1),\ j=0,\cdots,T$
		\STATE $iter \gets 0$
		\WHILE{$iter<10$}
		\STATE simulate to obtain $\mathbf{A}$
		\STATE $\mathbf{\alpha} \gets \mathbf{p}_{m}^{T}\mathbf{A}$
		\STATE $\epsilon \gets (\max{\mathbf{\alpha}}-\min{\mathbf{\alpha}})/{\mathbb E[\mathbf{\alpha}}]$
		\STATE $iter \gets iter+1$
		\IF{$\epsilon > 10^{-3}$}
		\STATE $\mathbf{p}_{m}^{T} \gets \mathbf{1}_{1\times (T+1)}\mathbf{A}^{\dagger}$
		\STATE $\mathbf{p_m} \gets \mathbf{p_m}/\sum_{j=0}^{T}p_j$
		\ELSE
		\STATE break
		\ENDIF
		\ENDWHILE
	\end{algorithmic}
}
\end{algorithm}

In Algorithm 1, the initial value of $\mathbf{p}_{m}$ is obtained by
uniformly selecting the multiplier $\bm g$ from the set $\mathbb{Z}%
_{2^{m}}\backslash \{0\}$. Note that this initial value can also be replaced
by other empirical values, which may speed up the convergence of the
algorithm. $\mathbf{\Theta }^{\dagger }$ denotes the Moore-Penrose inverse
of $\mathbf{\Theta }$. The updated value of $\mathbf{p}_{m}$ may be of
negative entry, which will be forced to 0. In this case, the $i$-th row of $%
\mathbf{\Theta }$ corresponding to $p_{i}=0$ are all zeros. The solution
obtained by Moore-Penrose inverse is the least square solution\cite%
{sawyer2006generalized}. When there is no $\mathbf{p}_{m}$ that meets $%
\mathbf{p}_{m}^{T}\mathbf{\Theta }=\alpha \mathbf{1}_{1\times (T+1)}$, the
obtained solution ensures that the difference among the entries of $\mathbf{p%
}_{m}^{T}\mathbf{\Theta }$ are minimized.

%
%
%

With the optimized irregular multiplier distribution, numerical results show
that the symmetric Gaussian approximation is maintained well. As such, the
optimization of the irregular node degree distribution with EXIT chart is in
line with the trajectory of mutual information in the iterative decoding
process.

\subsection{Computation of EXIT Functions}

Here we characterize input-output mutual information (MI), i.e. the EXIT
functions \cite{ten2004design}, of VN and CN of the D-IRA ring codes. The
utilization of random coset has the same effect as the output symmetry of
binary LDPC codes\cite{910577,910578,1226601,1542413}. The EXIT function of
VN with degree $d_{v}$ is given by\cite{ten2003design,ten2004design}
\begin{equation}
I_{E,VND}(I_{A},d_{v})=J((d_{v}-1)\cdot J^{-1}(I_{A}))  \label{E2}
\end{equation}%
where $J(\sigma ^{2})=I(C;\Lambda )$ denotes the MI between the genuine
codeword and the LLR sequence, which is characterized with a
single-parameter $\sigma ^{2}$. For CNs, if the distribution of the output
LLR vector satisfies
\begin{equation}
\mathrm{Pr}\left[ {\lambda }\mid C=\omega \right] =\mathrm{Pr}\left[ {%
\lambda }^{+\omega }\mid C=0\right] ,  \label{Eq18}
\end{equation}%
i.e., the symmetric condition is met, then the (normalized) mutual
information can be expressed as
\begin{equation}
I(C;\Lambda )=1-\mathbb{E}\left[ \log _{2^{m}}\left(
1+\sum_{i=1}^{2^{m}-1}e^{-\lambda _{i}}\right) \right] .
\end{equation}%
The operator $\mathbb{E}[\cdot ]$ denotes the expectation over the LLR
vectors, which can be approximated numerically in practice. In this paper,
we use the assumption in (\ref{Eq18}). This avoids multi-dimensional
integration in the calculation of mutual information and yield a
satisfactory performance as we will see later.

For edges with multiplier of zero-divisors, the output probabilities are
block-wisely repeated. If the zero-multiplier $\xM0$ of the zero divisor is $%
m_{j}$, the first ($m_{j}-1$) entries of $\mathbf{\Lambda }$ contains all
information, i.e.,
\begin{equation}
{\small \begin{aligned} I(C;{\bm \Lambda}_{1:m_j-1})&=1-\mathbb
E\left[\log_{2^m}\left(1+\sum_{i=1}^{m_j-1}e^{-w_i}\right)\right]=1-%
\mathbb
E\left[\log_{2^m}\left(\frac{2^m}{m_j}\left(1+\sum_{i=1}^{m_j-1}e^{-w_i}\right)%
\right)\right]\\ &=1-\mathbb
E\left[\log_{2^m}\left(1+\sum_{i=1}^{2^m-1}e^{-w_i}\right)\right], \end{aligned}}
\label{E2_3}
\end{equation}%
where the equality in the last step is due to the block-wise repetition of
probabilities.

For CNs with degree $d_{c}$, the output mutual information normalized by $%
\log _{2}2^{m}$ is obtained by
\begin{equation}
I_{E,CND}\left( I_{A};d_{c},\sigma _{z}^{2}\right) =1-\mathbb{E}\left[ \log
_{2^{m}}\left( 1+\sum_{i=1}^{2^{m}-1}e^{-\lambda _{i}}\right) \right]
\label{E2_4}
\end{equation}%
where the input LLR vectors obey the joint Gaussian distribution\cite{910580}
with parameter $\sigma ^{2}$ meeting $J(\sigma ^{2})=I_{A}$, the AWGN
variance is $\sigma _{z}^{2}$. For $I_{A}=1$, we have $%
\sum_{i=1}^{m_{j}-1}e^{-\lambda _{i}}=0$ and
\begin{equation}
I_{E,CND}\left( I_{A}=1;d_{c},\sigma _{z}^{2}\right) =\frac{1}{\log _{2}2^{m}%
}\sum_{j=0}^{T}p_{j}\log _{2}(m_{j})<1,  \label{E2_5}
\end{equation}%
according to the first equality in (\ref{E2_3}).

\subsection{Optimization of Irregular Node Degree Distribution}

The distribution of node degrees is defined by polynomials
\begin{equation}
\varphi (x)=\sum_{i=2}^{D_{v}}\varphi _{i}x^{i-1}\ \mathrm{and}\ \rho
(x)=\sum_{j=1}^{D_{c}}\rho _{j}x^{j-1},
\end{equation}%
where $D_{v}$ and $D_{c}$ denote the maximum degree of VNs and CNs
respectively, $\lambda _{i}$ denotes the fraction of edges connected to
variable nodes with degree $i$ and $\rho _{j}$ denotes that of check nodes
with degree $j$. The coding rate w.r.t. $(\varphi ,\rho )$ is
\begin{equation}
R_{s}=\frac{\sum_{i=2}^{D_{v}}\varphi _{i}/i}{\sum_{j=1}^{D_{c}}\rho _{j}/j}.
\label{E3_2}
\end{equation}

For given degree distribution $(\varphi ,\rho )$, the effective EXIT
functions are
\begin{equation}
\begin{aligned}
I_{E,VND}\left(I_A\right)&=\sum_{i=2}^{D_v}\lambda_{i}I_{E,VND}\left(I_A;i%
\right)\\
I_{E,CND}\left(I_A;\sigma_z^2\right)&=\sum_{j=1}^{D_c}\rho_{j}I_{E,CND}%
\left(I_A;j,\sigma_z^2\right). \end{aligned}
\end{equation}

For a given degree distribution of CNs, we utilize linear programming to
optimize the degree distribution of VNs. The constraints are (\ref{E3_2})
and
\begin{equation}
I_{E,VND}\left(I_A\right)>I_{E,CND}^{-1}\left(I_A;\sigma_z^2\right),\quad0%
\leq I<1.  \label{E3_4}
\end{equation}
For given degree distribution of VNs, the constraint (\ref{E3_4}) becomes
\begin{equation}
I_{E,CND}\left(I_A;\sigma_z^2\right)>I_{E,VND}^{-1}\left(I_A\right),\quad0%
\leq I<1.  \label{E3_5}
\end{equation}


The algorithm for optimizing the node degree distribution is shown in
\textbf{Algorithm 2}. {\small
\begin{algorithm}
	\caption{Solving the optimal node degree distribution}
	\label{alg2}
{\small
	\begin{algorithmic}
		\REQUIRE $R_s$, $I_{E,VND}\xLRs{I_A;d_v}$ and $I_{E,CND}\xLRs{I_A;d_c,\sigma_z^2}$
		\ENSURE expressions (\ref{E3_2}), (\ref{E3_4}), (\ref{E3_5}) with $I_{A,CND}\in\xLRm{0,0.85}$
		\STATE $\rho^{(0)}:\ (\rho_1,\rho_3) \gets (0.1,0.9)$
		\STATE $n \gets 1$
		\WHILE{$n<10$}
		\STATE
		use $\rho^{(n-1)}$, $I_{E,CND}\xLRs{I_A;d_c,\sigma_z^2}$ to get $I_{A,CND}\xLRs{I_E;\sigma_z^2}$
		\STATE $\lambda^{(n)}\gets \mathop{\arg\max}\limits_{\lambda}\xLRs{\min\xLRs{\frac{I_{E,VND}\xLRs{I_A}-I_{A,CND}\xLRs{I_E;\sigma_z^2}}{\sqrt{1+{I_{A,CND}^{'}\xLRs{I_E;\sigma_z^2}}^2}}}}$
		\STATE use $\lambda^{(n)}$, $I_{E,VND}\xLRs{I_A;d_v}$ to get $I_{A,VND}\xLRs{I_E}$
		\STATE $\rho^{(n)}\gets \mathop{\arg\max}\limits_{\rho}\xLRs{\min\xLRs{\frac{I_{E,CND}\xLRs{I_A;\sigma_z^2}-I_{A,VND}\xLRs{I_E}}{\sqrt{1+{I_{A,VND}^{'}\xLRs{I_E}}^2}}}}$
		\STATE $n \gets n+1$
		\ENDWHILE
	\end{algorithmic}
}
\end{algorithm}
} In Algorithm 2, $I_{A}\left(I_E\right)$ denotes the inverse function $%
I_{E}^{-1}\left(I_A\right)$ and $I_{A}^{^{\prime }}\left(I_E\right)$ denotes
its derivative. This algorithm is set to maximize the narrowest gap between
the two EXIT curves.

\subsection{Partially Random Interleaver}

Here we introduce the partially random interleaver of the D-IRA ring codes.
With the optimized multiplier distribution and node degree distribution, the
edges of CNs and the associated multipliers are determined. The interleaver
connects these edges to the VNs with a specific order\cite{1201066}\cite%
{UppalTcom13}. According to the degrees of CNs and VNs and the multipliers,
the edges are divided into two categories.

(\romannumeral1) For degree-2 and 3 VNs, the edges have multipliers of
regular elements (no zero-divisors).

(\romannumeral2) For VN with degree greater than 3, the edges are allowed to
have zero-divisors. The number of edges with zero-divisors cannot exceed a
certain proportion of its node degree (e.g., 1/3).

Then, the edges of VNs are randomly connected subject to the above
constraint, and hence the name \textquotedblleft partially
random\textquotedblright\ interleaver. By and large, the edges of
multipliers of zero-divisors are connected to VNs of large degrees. This can
effectively reduce the effect of the ambiguity due to the multipliers of
zero-divisors in the LLR calculation of the CNs.

\subsection{Check Nodes with Degree One are no Longer Required}

In existing IRA codes over Galois fields, a portion of $d_{c}=1$ CNs must be
used. For example in \cite{ten2003design}, 20\% of the CNs are forced to
have degree 1. This is due to that the input mutual information obtained
purely from the channel intrinsic information is zero for $d_{c}\geq 2$ CNs,
and the iterative decoding process cannot commence without degree-1 CNs.
This part will show that in D-IRA ring codes such requirement is relaxed. As
such, the tunnel between the EXIT curves may be better exploited.

\begin{lemma}
	Consider a CN with at least one input edge whose multiplier is a regular element. If the input vector w.r.t. the regular element has equal probability, the output vector also has equal probability.\\
	\begin{proof}
		When $g$ is a regular element, ${\bm x}^{\times g}$ in (\ref{E1_1}) is an arrangement of ${\bm x}$. Without losing generality, the $k$-th input edge is labeled with regular element and has equiprobable probability vector ${\bm r}^k$. So there is $\bar{\bm r}^k=\xLRm{\frac{1}{q},\cdots\,,\frac{1}{q}}$ and ${\rm DFT}(\bar{\bm r}^k)=\xLRm{1,0,\cdots\,,0}$. The first element of the DFT vector is the sum of the probability vector, which is 1 for all input edges. According to (\ref{E5}), the output is
		\begin{equation}
			\bar{\bm l}={\rm IDFT}\xLRs{\xLRm{1,0,\cdots\,,0}}=\xLRm{\frac{1}{q},\cdots\,,\frac{1}{q}}.
		\end{equation}
		Because every element in ${\bm l}$ is an element in $\bar{\bm l}$, the output probability vector is also equal probability.
	\end{proof}
	\label{lemma1}
\end{lemma}

Consider a CN with $d_{c}>1$. This CN has the
constraint equation $\sum_{n=1}^{d_{c}+2}h_{n}a_{n}=0$. Consider the initial
state, i.e. $I_{A,CND}=0$. That is, the initial probability vector ${\bm r}%
^{n}=\left[ \frac{1}{2^{m}},\cdots \,,\frac{1}{2^{m}}\right] $, $n=1,\cdots
,d_{c}$. If there is at least one regular element in the multipliers $%
h_{1},\cdots ,h_{d_{c}}$, the output probability vectors ${\bm l}^{d_{c}+1}=%
\left[ \frac{1}{2^{m}},\cdots \,,\frac{1}{2^{m}}\right] $ and ${\bm l}%
^{d_{c}+2}=\left[ \frac{1}{2^{m}},\cdots \,,\frac{1}{2^{m}}\right] $
according to Lemma \ref{lemma1}.
\begin{figure}[h]
\centering
\includegraphics[scale=0.4]{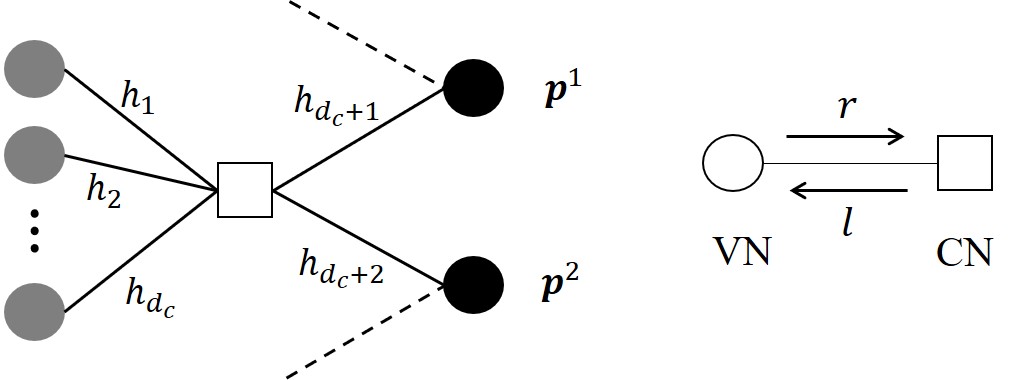}
\caption{Illustration of a CN of degree $d_{c}$ and the multipliers.}
\end{figure}

${\bm l}^{d_{c}+1}$ and ${\bm l}^{d_{c}+2}$ are also equiprobable for the
previous or next CN, so ${\bm r}^{d_{c}+1}$ and ${\bm r}^{d_{c}+2}$ are
equal to the probability vectors of the channel, which are ${\bm p}^{1}$ and
${\bm p}^{2}$ respectively. Let $\mathbf{Fp}$ denote the product of the a
posteriori probability vector from the channel
\begin{equation}
\mathbf{Fp}=\mathrm{DFT}(\bar{\bm p}^{1})\cdot \mathrm{DFT}(\bar{\bm p}^{2})
\end{equation}%
where $\bar{\bm p}^{1}$ and $\bar{\bm p}^{2}$ are the multiplication cycles
with $h_{d_{c}+1}^{-1}$ and $h_{d_{c}+2}^{-1}$ respectively.

Consider the output probability vectors of CN. If there are two or more
regular elements in the multipliers $h_1,\cdots,h_{d_c}$, all of the output
vectors are equal probability according to Lemma \ref{lemma1}, that is, $%
I_{E,CND}=0$. There is $I_{E,CND}>0$ only when the check node with degree $%
d_c$ has at least $d_c-1$ zero-divisors.

As multipliers of the same check node should avoid too many zero-divisors,
we consider the CN with degree 2 here. Suppose $h_{1}$ is a regular element
and $h_{2}$ is a zero divisor. ${\bm l}^{2}$ is equiprobable and
\begin{equation}
\begin{aligned} \bar{\bm l}^1&={\rm IDFT}\left({\rm DFT}(\bar{\bm r}^2)\cdot
\mathbf{Fp}\right). \end{aligned}
\end{equation}

The DFT vector of the multiplication cycle of the probability vector is dual
to the multiplication cycle of the DFT vector of the probability vector,
that is
\begin{equation}
{\small \begin{aligned} {\rm DFT}(\bar{\bm r}^2)&=\left({\rm DFT}({\bm
r}^2)\right)^{\times (-h_2)}=\left(\left[1,0,\cdots\,0\right]\right)^{\times
(-h_2)}\\ &=\begin{cases} 1 \quad i\otimes h_2=0 \\ 0 \quad i\otimes
h_2\neq0 \end{cases}. \end{aligned} }
\end{equation}
The zero-multiplier of $h_2$ is $\xM0(h_2)<q$. So at least two elements in
the DFT vector are 1.

According to (\ref{E2_4}), the mutual information corresponding to an edge
can be rewritten as
\begin{equation}
1-\log _{2^{m}}\left( 1+\sum_{i=1}^{q-1}e^{-\lambda _{i}}\right) =1+\log
_{2^{m}}\left( p_{0}\right)
\end{equation}%
That is, the output mutual information is only related to the zeroth element
of the probability vector. According to the IDFT calculation,
\begin{equation}
l_{0}^{1}=\bar{l}_{0}^{1}=\frac{1}{2^{m}}\sum_{i=0}^{2^{m}-1}\left[ \mathrm{%
DFT}(\bar{\bm l}^{1})\right] _{i}
\end{equation}

Therefore, the mutual information of the edge corresponding to ${\bm l}^{2}$
is 0 and that corresponding to ${\bm l}^{1}$ is
\begin{equation}
1+\log _{2^{m}}\left( l_{0}^{1}\right) =\log _{2^{m}}\left( \sum_{i=0}^{q/\xM%
0(h_{2})-1}\left[ \mathbf{Fp}\right] _{i\cdot \xM0(h_{2})}\right)
\end{equation}

The above expressions are the same for multipliers with the same $\xM0$, so
the corresponding MI is
\begin{equation}
I_{E}\left( m_{j},\sigma _{z}^{2}\right) =\mathbb{E}\left[ \log
_{2^{m}}\left( \sum_{i=0}^{2^{m}/m_{j}-1}\left[ \mathbf{Fp}\right] _{i\cdot
m_{j}}\right) \right]
\end{equation}%
where $m_{0}=2^{m}$ and $I_{E}\left( m_{0},\sigma _{z}^{2}\right) =0$
include the case that $h_{2}$ is a regular element.


For CNs with $d_{c}=2$, if there is at least one regular element, the
initial output MI is
\begin{equation}
I_{E,CND}\left( I_{A}=0;d_{c}=2,\sigma _{z}^{2}\right)
=\sum_{j=1}^{T}p_{j}I_{e}\left( m_{j},\sigma _{z}^{2}\right) .
\end{equation}%
If the multiplier takes more zero-divisors (especially the zero-divisors
with smaller $\xM0$), the initial output mutual information of the CN with
degree 2 is non-zero which enables the commencement of iterative decoding.
Therefore, CNs with degree 1 is no longer required. We conjecture that the
performance improvement of D-IRA ring codes (to be shown in the numerical
result section) is due to the relaxation of the requirement of degree-1 CNs.

\section{D-IRA Ring Codes for Multi-user Networks}


We now turn to our real interest: to exploit structured binning with D-IRA
ring codes in multi-user networks, with practical $2^{m}$-PAM signaling.
Here we only present the treatment for the CF and DPC setups, which can be
extended to other network configurations. The techniques presented in this
section do not apply for conventional BICM, SCM, TCM and existing GF$\left(
2^{m}\right) $ modulation codes.

\subsection{D-IRA Ring codes for Compute-forward}

Let $\mathbf{w}_{1}$,$\cdots ,\mathbf{w}_{K}$ denote the $2^{m}$-ary message
sequences of the $K$ users. Each user's message sequence is encoded as in (%
\ref{Eq_encodinggeneral}) via a common D-IRA ring code, yielding coded
sequences $\mathbf{c}_{1}$,$\cdots ,\mathbf{c}_{K}$. Let $\mathbf{x}_{1}$,$%
\cdots ,\mathbf{x}_{K}$ denote the resultant $2^{m}$-PAM sequences after the
one-to-one mapping in (\ref{Eq_PAMmapping}), which are transmitted
simultaneously. The received signal is
\begin{equation}
\mathbf{y}=\sum_{i=1}^{K}h_{i}\mathbf{x}_{i}+\mathbf{z.}
\end{equation}%
where $h_{i}$ denotes the channel gain of user $i$. The receiver aims to
compute $L$ linear message combinations
\begin{equation}
\mathbf{u}_{l}\triangleq \alpha _{l,1}\mathbf{w}_{1}\oplus \alpha _{l,2}%
\mathbf{w}_{2}\oplus \cdots \oplus \alpha _{l,K}\mathbf{w}_{K},l=1,\cdots ,L
\label{Eq_LinearMessageCombination}
\end{equation}%
where $\alpha _{l,1},\cdots ,\alpha _{l,K}\in \mathbb{Z}_{2^{m}}$ are the
coefficients. Let $\mathbf{\alpha }_{l}=\left[ \alpha _{l,1},\cdots ,\alpha
_{l,K}\right] ^{T}$ be a \emph{coefficient vector}, and let $\mathbf{A}=%
\left[ \mathbf{\alpha }_{1},\cdots ,\mathbf{\alpha }_{L}\right] ^{T}$ be the
\emph{coefficient matrix}.

We note that CF or linear PNC is not only confined to the toy example of
two-way relay channel studied in the early stage of this subject. It is now
understood that CF or linear PNC applies to a wide range of multi-user
communication configurations, such as multiple-access (MA), broadcast (BC),
distributed MIMO, integer-forcing (IF) MIMO detection, IF precoding,
interference alignment, multi-access relay and etc., with remarkable
advances in achievable rates \cite%
{YangTWC17_NOMA,NtranosISIT13,nazer2011compute,ZhanIT14}. As such, we
consider the above model that consists of $K$ users and one receiver, where
the receiver is set to compute $L$ linear message combinations. Such a model
is the core to any systems operated with CF. For the case of $K=2$, $L=1$,
the model can be used to represented the uplink phase of a two-way relay
channel. For the case of $L=K$, when coefficient matrix $\mathbf{A}$ is
invertible over $\mathbb{Z}_{2^{m}}$, the model can be used to represent an
uplink MA system. The subsequent operations, such as the downlink phase of
the two-way relay channel and the multiplication of the inverse of the
channel coefficient matrix in MA and distributed MIMO, are diverse and thus
not presented in this paper. Interested readers may find the details in \cite%
{yang2015achieving,YangTWC17_NOMA}.

\subsubsection{Structured Binning}

Here, a certain $\mathbf{u}_{l}$ specifies a set of candidates $\left[
\mathbf{w}_{1},\cdots ,\mathbf{w}_{K}\right] $ of the same underlying linear
message combination, which is essentially a \textquotedblleft \emph{bin-index%
}\textquotedblright\ in the jargon of network information theory. In
particular, the bin-index $\mathbf{u}_{l}$ is given by a linear structure
specified in (\ref{Eq_LinearMessageCombination}), hence the name
\textquotedblleft \emph{structured binning\textquotedblright }. Apparantly,
different choice of coefficient vectors yield different binning structures.
Roughly speaking, the $L$ coefficient vectors with the \textquotedblleft
best\textquotedblright\ binning structures, such that the bin-indices can be
most reliably computed, should be selected and utilized.

Traditional non-PNC schemes completely decodes $\mathbf{w}_{1}$,$\cdots ,%
\mathbf{w}_{K}$ individually, and then form $\mathbf{u}_{l}$. These schemes
are based on the notion of \emph{random coding} for the multi-user
communication. With lattice codes or ring codes, PNC can exploit structured
binning to directly computes $\mathbf{u}_{l}$ without the complete decoding,
yielding significant coding or even multiplexing gain. CF characterizes the
achievable rate for reliably computing of $\mathbf{u}_{l}$ based on the
existence of good lattice codes for structured binning. The implementation
of a D-IRA ring coded PNC system is depicted as follows.

Let $\mathbf{C=}\left[ \mathbf{c}_{1},\cdots ,\mathbf{c}_{K}\right] ^{T}$
stacks up all users' coded sequences generated by the D-IRA ring code.
Define
\begin{equation}
\mathbf{v}_{l}^{T}\triangleq \func{mod}\left( \sum_{i=1}^{K}a_{l,i}\mathbf{c}%
_{i}^{T},q\right) =\mathbf{a}_{l}^{T}\mathbf{\otimes }_{q}\mathbf{C}
\end{equation}%
as the $l$-th \textquotedblleft \emph{linear coded-sequence combination}".

\begin{property}
With the generator matrix $\mathbf{G}$ in (\ref{Eq_encodinggeneral}), we
have
\begin{eqnarray}
\mathbf{v}_{l} &=&\func{mod}\left( \sum_{i=1}^{K}a_{l,i}\mathbf{G\otimes }%
_{q}\mathbf{b}_{i},q\right) =\mathbf{G\otimes }_{q}\func{mod}\left(
\sum_{i=1}^{K}a_{l,i}\mathbf{b}_{i},q\right)  \notag \\
&=&\mathbf{G\otimes }_{q}\mathbf{u}_{l}.
\end{eqnarray}%
That is, a linear coded-sequence combination $\mathbf{v}_{l}$ and a linear
message combination $\mathbf{u}_{l}$ are also related by the multiplication
of $\mathbf{G}$ modulo-$q$.
\end{property}

Property 2 allows for: 1) calculating the symbol-wise APPs of $\mathbf{v}%
_{l} $ over the extended constellation, to be detailed momentarily; 2)
forward the resultant APP sequence to a decoder to compute $\mathbf{u}_{l}$.
We note that such treatment is impossible for non-lattice code based schemes
where Properties 2 does not hold. The implementation of 1) and 2) are
illustrated below:

Let $v_{l}\left[ t\right] $ and $y\left[ t\right] $ denote the $t$-th column
of $\mathbf{v}_{l}$ and $\mathbf{y}$, respectively. The receiver calculates
the symbol-wise APPs of the linear coded-sequence combinations $p\left( v_{l}%
\left[ t\right] |y\left[ t\right] \right) $. This can be implemented in
parallel or in succession for $l=1,\cdots ,L$. Using the Baye's rule, we
obtain%
\begin{equation}
p\left( v_{l}\left[ t\right] =\omega |\mathbf{y}\left[ t\right] \right)
=\sum _{\substack{ x_{1}\left[ t\right] ,\cdots ,x_{K}\left[ t\right] :\text{
}  \\ \mathbf{\alpha }_{l}^{T}\mathbf{\otimes c}\left[ t\right] =\omega }}%
p\left( y\left[ t\right] |\sum_{i=1}^{K}h_{i}x_{i}\left[ t\right] \right)
,\omega =0,\cdots ,2^{m}-1.  \label{Eq_symbolwiseAPP_1stedition}
\end{equation}%
It equals to the sum of the likelihood functions of the $2^{m}$ candidates
in the bin with index $\omega$. The APPs are forwarded to $L$ D-IRA ring
code BP decoders, which yield decision on $\mathbf{u}_{l}$, $l=1,\cdots ,L$,
in the case with parallel computing. More details, such as successive
computing, can be found in \cite{yang2015achieving}, \cite{YangTWC17}.

\subsubsection{D-IRA Ring Code Optimization}

The optimized design of the D-IRA ring codes for this CF setting needs to
evaluate the input-output mutual information transfer function that takes
into account the symbol-wise APP calculation over the extended constellation
depicted above. In particular, such operation result in a varied EXIT
function for the combined check-accumulator. Thus, the curve-fitting of the
EXIT functions need to adapt to these change in generating the optimized
D-IRA ring code degree-profile. For the CF setting, it is empirically found
that a good single-user D-IRA ring code tends to be a good code for the $K$%
-user CF setting as well, as we will shown in the numerical result section.

\remark The zero-divisors can also be used as coefficients in certain
setups. For MA where $L=K$, it can be shown that as long as $\func{mod}%
\left( |\mathbf{A}|,q\right) $ is a regular element, all users' message can
be recovered.

\subsection{D-IRA Ring codes for Linear Dirty Paper Coding}

Consider the DPC setting where a base station (BS) wants to deliver a
message to a user equipment (UE), subject to some interference at the UE
receiver \cite{CostaTIT83}. Let $\mathbf{w}$ denote the $2^{m}$-ary message
sequence and $\mathbf{x}$ denote the coded $2^{m}$-PAM signal sequence
transmitted by BS. The UE receives%
\begin{equation}
\mathbf{y}=\mathbf{x}+\mathbf{s}+\mathbf{z}  \label{Eq Channel Model}
\end{equation}%
where $\mathbf{s}$ is the interference and $\mathbf{z}$ is the additive
white Gaussian noise (AWGN) of mean zero and variance $\sigma ^{2}$. The
interference $\mathbf{s}$ is known by the BS transmitter but not by the UE
receiver. For applications in the downlink of cellular network, $\mathbf{s}$
could be the signal to another UE. It is well-known that DPC is required to
achieve the capacity of this multi-source channel.

For the clarity of presentation, let us omit the power normalization factor $%
\gamma $ in (\ref{Eq_PAMmapping}). For integer interference, i.e., $\mathbf{%
s\in }$ $%
\mathbb{Z}
^{n}$, a simple linear DPC method is given by \cite{YangICC22}
\begin{equation}
\mathbf{x}\mathbf{=c\ominus \mathbf{s}-}\frac{q-1}{2}
\label{Eq DPC Encoding}
\end{equation}%
where $a\mathbf{\ominus }b=\func{mod}\left( a-b,2^{m}\right) $. Integer
interference arises when $\mathbf{s}$ is a PAM signal from another UE. The
treatment for non-integer $\mathbf{s}$ can be found in \cite{YangICC22}, and
is not presented in this paper due to space limitation.


The UE receiver does not know $\mathbf{s}$, but has knowledge on the
statistics, e.g. the p.m.f. or p.d.f. of $\mathbf{s}$. Let
\begin{equation}
\mathbf{r}=\mathbf{x}+\mathbf{s}
\end{equation}%
denote the signal plus interference (without noise) at UE receiver. With (%
\ref{Eq Channel Model}) and (\ref{Eq DPC Encoding}), we obtain%
\begin{equation}
\mathbf{r=\mathbf{c}\ominus s}_{Q}+\mathbf{s}_{Q}\mathbf{-}\frac{q-1}{2}
\label{Eq_SignalPlusInterference}
\end{equation}%
which is guaranteed to belong to a extended codebook of the ring code \cite%
{YangICC22}. This does not hold for the GF($2^{m}$) codes \cite%
{chiu2009bandwidth}. The receiver is set to compute the bin-index with the
following procedures:

1) Upon receiving $\mathbf{y}$ in (\ref{Eq Channel Model}), compute the
symbol-wise APPs w.r.t. the entries of the coded sequence $\mathbf{c}$. For
the $t$th symbol, $t=1,\cdots ,n$, the APP for $c\left[ t\right] =i$, $%
i=0,\cdots ,2^{m}-1$, is%
\begin{eqnarray}
p\left( c\left[ t\right] =i|y\left[ t\right] \right) &=&\dsum\limits_{r\left[
t\right] \in \overline{\mathbb{C}}\left( i\right) }p\left( r\left[ t\right]
|y\left[ t\right] \right) \propto \dsum\limits_{r\left[ t\right] \in
\overline{\mathbb{C}}\left( i\right) }p\left( y\left[ t\right] |r\left[ t%
\right] \right) p\left( r\left[ t\right] \right)  \notag \\
&=&\frac{1}{\beta }\dsum\limits_{r\left[ t\right] \in \overline{\mathbb{C}}%
\left( i\right) }\exp \left( -\frac{\left( y\left[ t\right] -r\left[ t\right]
\right) ^{2}}{2\sigma ^{2}}\right) p\left( r\left[ t\right] \right)
\label{Eq_DecodingSymbolwiseAPP}
\end{eqnarray}%
where $\beta $ is a normalization factor to ensure $\dsum\limits_{i=0,\cdots
,q-1}p\left( c\left[ t\right] =i|y\left[ t\right] \right) =1$, $\overline{%
\mathbb{C}}\left( i\right) $ denotes the extended constellation of $2^{m}$%
-PAM. The a priori probability of $p\left( r\left[ t\right] \right) $ is
obtained from the statistics of $\mathbf{s}$ \cite{YangICC22}.

2) The APPs are forwarded to the D-IRA BP decoder, which yields the decoding
output $\widehat{\mathbf{w}}$.


The optimized design of D-IRA ring codes for DPC utilizes the input-output
MI transfer function that takes into account the DPC channel and symbol-wise
APP soft detector. Then, the optimization for the irregular-degree
multiplier and irregular node degree distributions follows the procedures in
Sections III and IV, Algorithms I and II. Our optimized solution for the
D-IRA ring coded DPC is given in Table. \ref{Table_Degree_Distribution}.

\begin{table}[h]
\caption{Optimized Degree Profile of D-IRA ring codes for DPC.}
\label{Table_Degree_Distribution}\centering
{\center
{\tiny
\begin{tabular}{|c|c|c|c|}
\hline
& Repeat Node Degree & Combiner Node Degree & Distribution of Multipliers \\
\hline
$q$=4 Rate=1/2 & $%
\begin{array}{c}
0.0008x^{2}\text{+}0.779x^{3}\text{+}0.0718x^{9} \\
\text{+}0.1097x^{12}\text{+}0.0256x^{20}\text{+}0.0131x^{60}%
\end{array}%
$ & $%
\begin{array}{c}
0.084x\text{+}0.0344x^{2} \\
\text{+}0.8804x^{3}\text{+}0.0012x^{5}%
\end{array}%
$ & $\left[ p_{1,}p_{2},p_{3}\right] $=$\left[ 0.4856,0.1657,0.3487\right] $
\\ \hline
$q$=8 Rate=1/2 & $%
\begin{array}{c}
0.3646x^{2}\text{+}0.3927x^{3}\text{+}0.019x^{6}\text{+}0.0064x^{8} \\
\text{+}0.1674x^{10}\text{+}0.0292x^{28}\text{+}0.0207x^{50}%
\end{array}%
$ & $%
\begin{array}{c}
0.084x\text{+}0.0991x^{2} \\
\text{+}0.7499x^{3}\text{+}0.0671x^{4}%
\end{array}%
$ & $%
\begin{array}{c}
\left[ p_{1,}\cdots ,p_{7}\right] \text{=}[0.2032,0.0882,0.2032 \\
0.0107,0.2032,0.0882,0.2032]%
\end{array}%
$ \\ \hline
$q$=8 Rate=2/3 & $%
\begin{array}{c}
0.4260x^{2}\text{+}0.3915x^{3}\text{+}0.0395x^{6}\text{+}0.0898x^{9} \\
\text{+}0.0318x^{12}\text{+}0.0171x^{32}\text{+}0.0043x^{60}%
\end{array}%
$ & $%
\begin{array}{c}
0.0284x\text{+}0.2016x^{2} \\
\text{+}0.7208x^{3}\text{+}0.0492x^{5}%
\end{array}%
$ & $%
\begin{array}{c}
\left[ p_{1,}\cdots ,p_{7}\right] \text{=}[0.2226,0.0467,0.2226 \\
0.0162,0.2226,0.0467,0.2226]%
\end{array}%
$ \\ \hline
$q$=16 Rate=5/8 & $%
\begin{array}{c}
0.6402x^{2}\text{+}0.0888x^{3}\text{+}0.0626x^{4}\text{+}0.1175x^{6} \\
\text{+}0.0481x^{12}\text{+}0.0257x^{21}\text{+}0.017x^{60}%
\end{array}%
$ & $%
\begin{array}{c}
0.0401x\text{+}0.061x^{2} \\
\text{+}0.0858x^{3}\text{+}0.012x^{4}%
\end{array}%
$ & $%
\begin{array}{c}
\left[ p_{1,}\cdots ,p_{15}\right] \text{=}[0.1196,0.0071,0.1196, \\
0.0053,0.1196,0.0071,0.1196, \\
0.0041,0.1196,0.0071,0.1196, \\
0.0053,0.1196,0.0071,0.1196,]%
\end{array}%
$ \\ \hline
\end{tabular}
} }
\end{table}

We note that the ring coded linear DPC generally outperforms binary coded
DPC. To be specific, ring coded linear DPC follows the notion of vector
quantization over lattices, and can be shown to minimize the quantization
error as $q$ increases. In contrast, a binary coded DPC is implemented based
on the soft probabilities w.r.t. the binary codes by dealing with the
many-to-one constellation mapping, which give rises to a performance gap to
the ring coded DPC, as will be shown in the numerical result section.

%

\section{Design Examples and Numerical Results}

\subsection{D-IRA Ring Codes for Point-to-point AWGN Channel}

This part presents the optimized code profile of D-IRA ring codes for AWGN
channel. For $q=4$, the coding rates under consideration are $%
R_{c}=1/4,1/2,3/4$, where the spectral efficiency are $R=1/2,1,3/2$
respectively. For $q=8$, the coding rates under consideration are $%
R_{c}=1/3,1/2,2/3$, where the spectral efficiency are $R=1,3/2,2$
respectively. The capacity limits w.r.t. these rates are obtained by
evaluating the mutual information with $4$-PAM and 8-PAM channel inputs.

\begin{table}[h]
\caption{The multiplier distribution for $q=4$.}
\label{table 4}\centering
\renewcommand{\multirowsetup}{\centering} \renewcommand{\arraystretch}{1.2}
{\small
\begin{tabular}{c|cc|cc|cc}
\hline
& \multicolumn{2}{c|}{R=0.5} & \multicolumn{2}{c|}{R=1.0} &
\multicolumn{2}{c}{R=1.5} \\ \hline
1 & 0.8420 & 0.1580 & 0.7965 & 0.2035 & 0.7223 & 0.2777 \\
2 & 0.8222 & 0.1778 & 0.8115 & 0.1885 & 0.8507 & 0.1493 \\
3 & 0.8022 & 0.1978 & 0.8004 & 0.1996 & 0.8610 & 0.1390 \\
4 & 0.7885 & 0.2115 & 0.7921 & 0.2079 & 0.8661 & 0.1339 \\
5 & 0.7773 & 0.2227 & 0.7855 & 0.2145 & 0.8693 & 0.1307 \\
6 & 0.7693 & 0.2307 & 0.7799 & 0.2201 & 0.8715 & 0.1285 \\ \hline
$\Omega$ & $\{1,3\}$ & $\{2\}$ & $\{1,3\}$ & $\{2\}$ & $\{1,3\}$ & $\{2\}$
\\ \hline
\end{tabular}%
}
\end{table}

\begin{table}[h]
\caption{The node degree distribution for $q=4$.}
\label{table 6}\centering\renewcommand{\multirowsetup}{\centering} %
\renewcommand{\arraystretch}{1.2} {\small
\begin{tabular}{c|c}
\hline
\multirow{2}[0]{*}{R=0.5} & $\varphi
(x)=0.1611x^{3}+0.0402x^{9}+0.1910x^{11}+0.1104x^{12}+0.4877x^{47}+0.0096x^{49}
$ \\
& $\rho (x)=0.0367x^{1}+0.5490x^{2}+0.0285x^{5}+0.3858x^{6}$ \\ \hline
\multirow{2}[0]{*}{R=1.0} & $\varphi
(x)=0.0800x^{2}+0.1492x^{3}+0.2379x^{4}+0.0101x^{9}+0.2384x^{10}+0.1657x^{13}+0.1187x^{22}
$ \\
& $\rho (x)=0.0080x^{1}+0.5012x^{2}+0.2532x^{3}+0.0239x^{4}+0.2136x^{6}$ \\
\hline
\multirow{2}[0]{*}{R=1.5} & $\varphi
(x)=0.2187x^{2}+0.3363x^{3}+0.1576x^{4}+0.0692x^{9}+0.1605x^{10}+0.0363x^{14}+0.0214x^{17}
$ \\
& $\rho (x)=0.3658x^{2}+0.5649x^{3}+0.0223x^{4}+0.0470x^{6}$ \\ \hline
\end{tabular}%
}
\end{table}

The optimized irregular multiplier distribution for $q=4$ is shown in TABLE %
\ref{table 4}, which is obtained by utilizing Algorithm 1. For each rate,
the portion of multipliers with coefficients \{1,3\} is given by the column
on the left, while that with coefficients \{2\} is given by the column on
the right. Each row denotes the portions of multipliers for a specific CN
degree, where the maximum CN degree is set to 6. For example, for the
spectral efficiency of 1 bits/symbol per real-dimension, for CN of degree 2,
18.85\% of the multipliers are \{2\} while 81.15\% of the multipliers are
either \{1\} or \{3\}. The multipliers with the same $M_{0}$ have identical
portion, e.g., multipliers \{1\} has portion 40.575\% while multipliers
\{3\} has portion 40.575\%. We note that certain amount of multiplier with
zeros-divisor \{2\} helps with the convergence behavior and the decoding
performance, relative to that without using a zero-divisor in the
multipliers. The ambiguity brought about by the zero-divisor multiplier can
be addressed in the BP decoding with the help from other nodes of different
multipliers and the partially random interleaver. As the rate increases,
less portion of multipliers with zero-divisor \{2\} is allocated. This can
be explained by considering that as the coding rate approaches 1, there will
be no redundancy digits and hence the zero-divisor \{2\} incurs ambiguity in
the division which cannot be solved. The node degree distribution is shown
in Table. \ref{table 6}. The polynomial $\lambda $ denotes the degree
distribution of the information (or repetition) nodes, while polynomial $%
\rho $ denotes the degree distribution of the CNs. The solution is obtained
by utilizing Algorithm 2.

{\small
\begin{table}[h]
\caption{The multiplier distribution for $q=8$.}
\label{table 5}{\small \centering
\centering
\begin{tabular}{c|ccc|ccc|ccc}
\hline
& \multicolumn{3}{c|}{R=1.0} & \multicolumn{3}{c|}{R=1.5} &
\multicolumn{3}{c}{R=2.0} \\ \hline
1 & 0.7506 & 0.2190 & 0.0304 & 0.6796 & 0.2571 & 0.0633 & 0.6971 & 0.2156 &
0.0872 \\
2 & 0.7549 & 0.2122 & 0.0329 & 0.7857 & 0.1734 & 0.0409 & 0.8700 & 0.0000 &
0.1300 \\
3 & 0.7336 & 0.2191 & 0.0473 & 0.7871 & 0.1573 & 0.0556 & 0.8684 & 0.0000 &
0.1316 \\
4 & 0.7153 & 0.2240 & 0.0607 & 0.7840 & 0.1463 & 0.0696 & 0.8642 & 0.0000 &
0.1358 \\
5 & 0.7007 & 0.2275 & 0.0718 & 0.7803 & 0.1380 & 0.0816 & 0.8592 & 0.0000 &
0.1408 \\
6 & 0.6893 & 0.2303 & 0.0804 & 0.7776 & 0.1301 & 0.0924 & 0.8546 & 0.0000 &
0.1454 \\ \hline
$\Omega$ & $\{1,3,5,7\}$ & $\{2,6\}$ & $\{4\}$ & $\{1,3,5,7\}$ & $\{2,6\}$ &
$\{4\}$ & $\{1,3,5,7\}$ & $\{2,6\}$ & $\{4\}$ \\ \hline
\end{tabular}
}
\end{table}
} {\small
\begin{table}[h]
\caption{The node degree distribution for $q=8$.}
\label{table 7}{\small \centering\centering
\centering%
\begin{tabular}{c|c}
\hline
\multirow{2}[0]{*}{R=1.0} & $\varphi
(x)=0.0600x^{2}+0.1064x^{3}+0.1113x^{5}+0.1442x^{8}+0.1442x^{13}+0.0915x^{25}+0.0248x^{26}+0.3175x^{57}
$ \\
& $\rho (x)=0.0175x^{1}+0.5990x^{2}+0.0054x^{5}+0.3781x^{6}$ \\ \hline
\multirow{2}[0]{*}{R=1.5} & $\varphi
(x)=0.0955x^{2}+0.2464x^{3}+0.2510x^{7}+0.0615x^{8}+0.0824x^{10}+0.1584x^{24}+0.1047x^{26}
$ \\
& $\rho (x)=0.0180x^{1}+0.5000x^{2}+0.2161x^{3}+0.0078x^{5}+0.2580x^{6}$ \\
\hline
\multirow{2}[0]{*}{R=2.0} & $\varphi
(x)=0.2103x^{2}+0.1852x^{3}+0.3199x^{6}+0.1124x^{17}+0.0746x^{21}+0.0976x^{48}
$ \\
& $\rho (x)=0.0004x^{1}+0.1500x^{2}+0.7856x^{3}+0.0078x^{4}+0.0561x^{6}$ \\
\hline
\end{tabular}
}
\end{table}
} The optimized irregular multiplier distribution for $q=8$ is shown in
TABLE \ref{table 5}. For each rate in the design, the portion of multipliers
with coefficients \{1,3,5,7\} is given by the column on the left, while that
with coefficients \{2,6\} is given by the column in the middle, that those
with \{4\} is given by the column on the right. In this example, there are
two types of zero-divisors: \{2,6\} and \{4\}. For a relatively high rate,
e.g., spectral efficiency of 2 bits/symbol per real-dimension, the portion
of multipliers with zero-multiplier coefficient \{2,6\} becomes zero except
for $d_{c}=1$. For a relatively low rate, the portion of multipliers with
zero-multiplier coefficient \{2,6\} is significantly greater than that for
\{4\}.
The irregular node degree distribution obtained from Algorithm 2 is shown in
Table. \ref{table 7}.

\begin{figure}[h]
\includegraphics[scale=0.20]{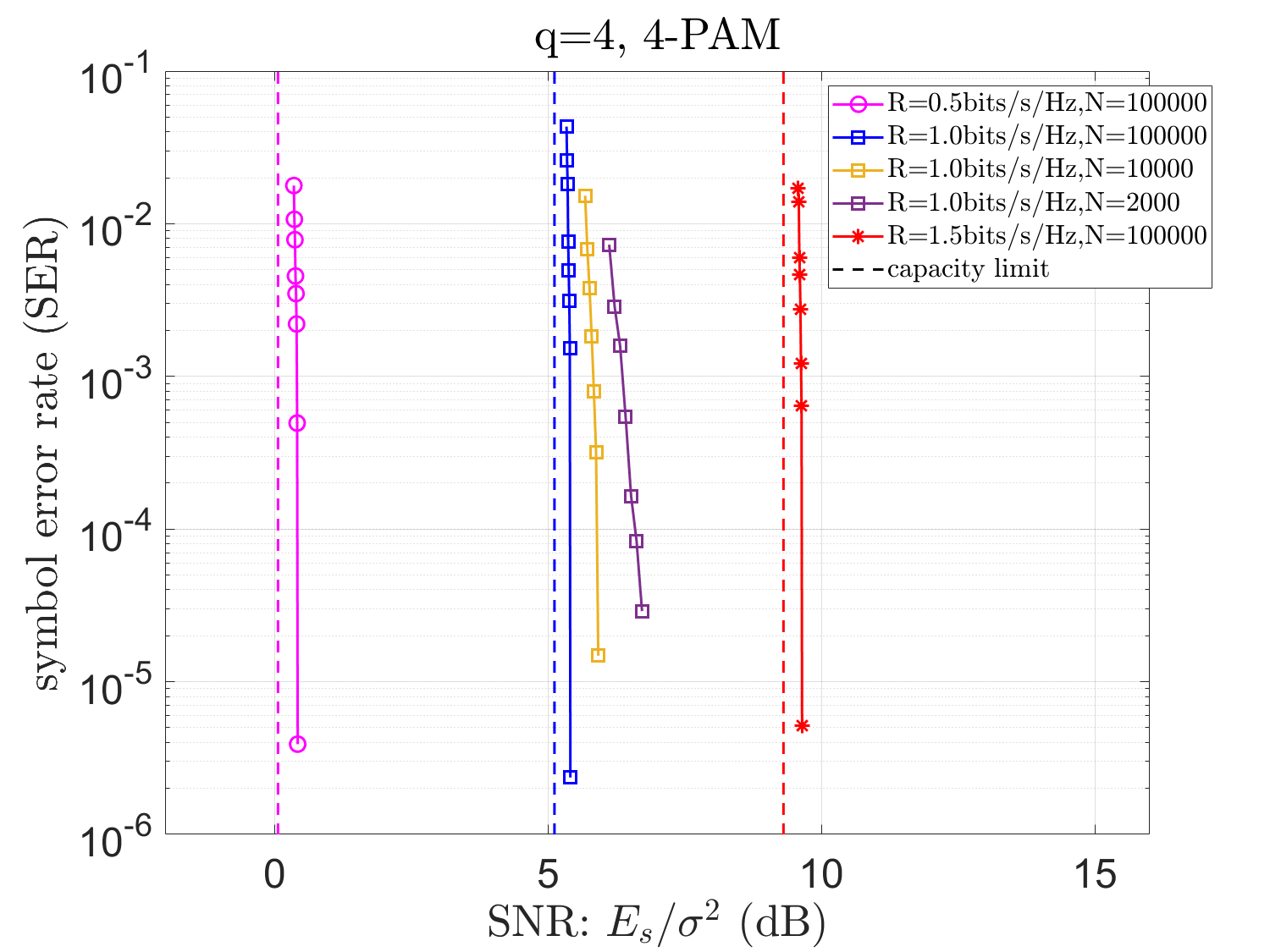}\centering
\caption{Performance of $q=4$ D-IRA ring codes with 4-PAM and various coding
rates.}
\label{Fig_BERq4}
\end{figure}

\begin{figure}[h]
\centering
\includegraphics[scale=0.20]{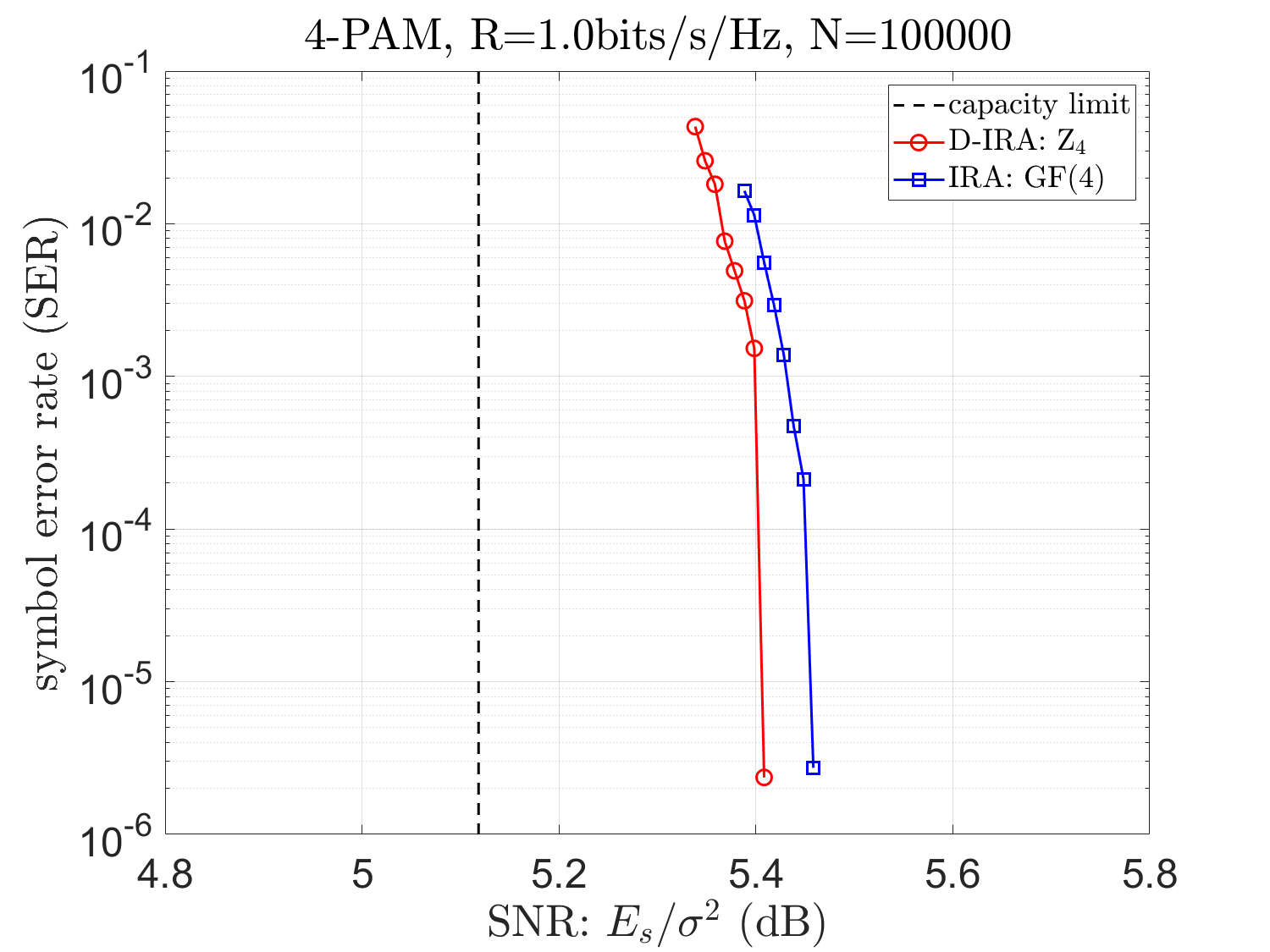}
\caption{Performance comparison of D-IRA ring codes and GF(4) IRA modulation
codes with 4-PAM.}
\label{Fig_BERGF4}
\end{figure}

\begin{figure}[h]
\centering
\includegraphics[scale=0.20]{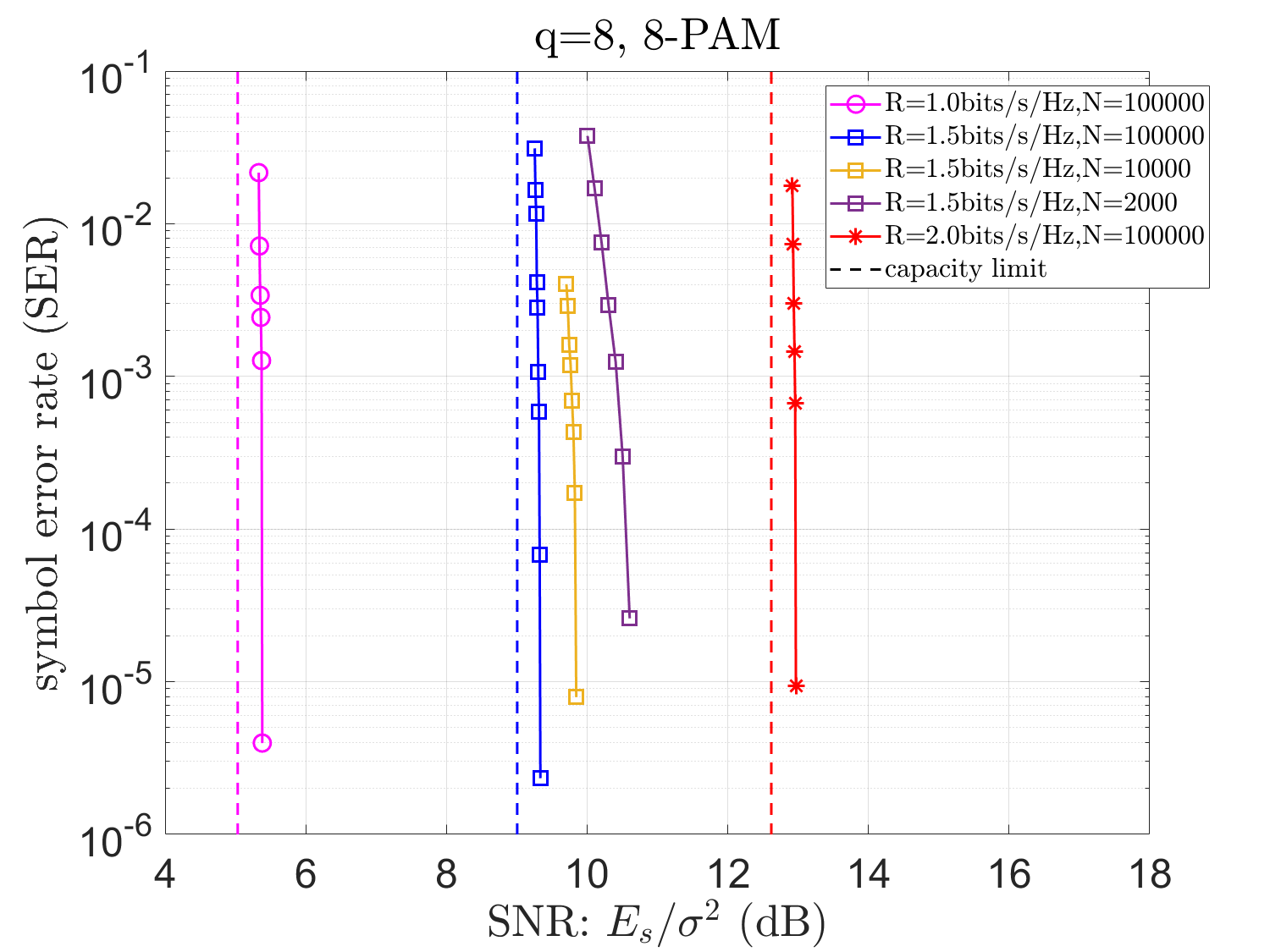}
\caption{Performance of $q=8$ D-IRA ring codes with 8-PAM and various coding
rates.}
\label{Fig_BERq8}
\end{figure}

\begin{figure}[h]
\centering
\includegraphics[scale=0.20]{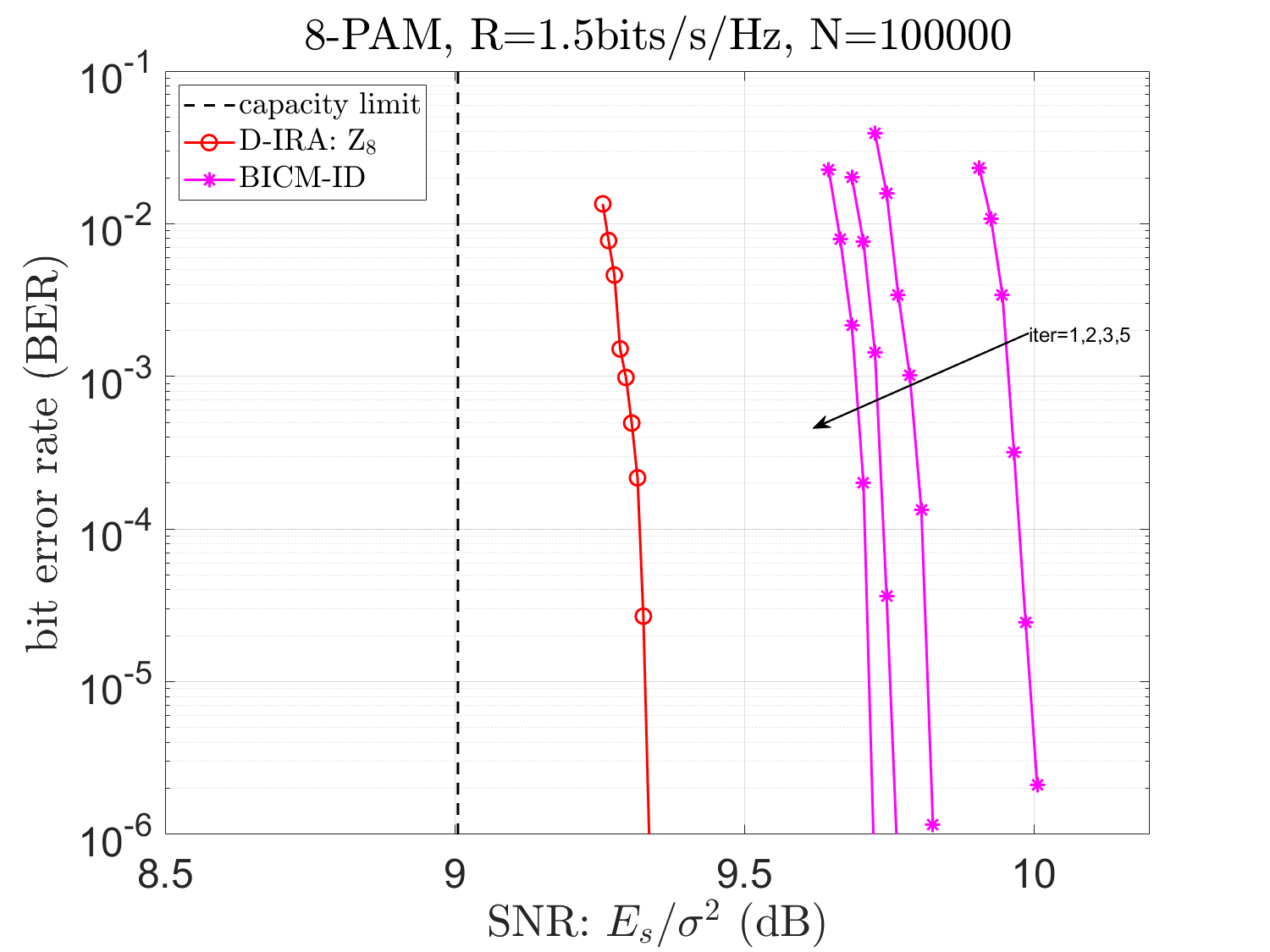}
\caption{Performance comparison of D-IRA ring codes and Gray mapping BICM-ID
with different iterations with 8-PAM.}
\label{Fig_BERBICM8}
\end{figure}

Fig. \ref{Fig_BERq4} shows the error-rate performance of the optimized D-IRA
ring coded $q$-PAM scheme of $q=4$, with the designed code profiles given in
the above. The codeword length is set to $n=100000$. At SER of $10^{-5}$,
the gaps between the symbol error performance of the optimized D-IRA ring
code and the capacity limits of 4-PAM are only 0.36, 0.29, 0.34 dB for
spectral efficiencies of $R=1/2,1,3/2$, respectively. In Fig. \ref%
{Fig_BERGF4}, we plot the error rate performance of the modulation code
based on GF(4) with optimized degree profile reported in \cite%
{chiu2009bandwidth}. It is observed that the proposed D-IRA ring code,
optimized via the pragmatic algorithms, exhibits a performance advantage of
0.05 dB at spectral efficiency of $R=1$. We conjecture that this is
primarily due to the fact that the existing binary and $q$-ary codes over
Galois fields require a portion of $d_{c}=1$ check nodes, otherwise the
iterative decoding will not commence. In contrast, in the proposed D-IRA
ring codes, even without $d_{c}=1$ CNs, the iterative process can still
start, owing to the existence of multipliers of zero-divisors. This may give
rise to further narrowed gap between the EXIT curves. At the current stage,
there is still no rigorous proof for this performance advantage.

Fig. \ref{Fig_BERq8} shows the performance of D-IRA ring codes for $m=3$ (or
$q=8)$. At SER of $10^{-5}$, the gaps between the optimized D-IRA modulation
code and the capacity limits of 8-PAM are only 0.35, 0.33, 0.35 dB for $%
R=1,3/2,2$, respectively. We also compare to existing BICM-ID scheme with
three levels of binary codes in Fig. \ref{Fig_BERBICM8}. It is observed that
the proposed D-IRA ring code yields 0.7 dB and 0.4 dB performance advantages
over the competing scheme with 1 and 5 BICM-ID iterations, respectively.

Fig. \ref{Fig_BERq4} and Fig. \ref{Fig_BERq8} also show the performance of
D-IRA ring codes for $q=4,R=1.0$ and $q=8,R=1.5$ with different codeword
lengths. The ring codes with different lengths adopt the same multiplier
distribution and node degree distribution according to TABLE \ref{table 4}-%
\ref{table 7}. At SER of $10^{-5}$, the gaps between the optimized D-IRA
modulation code and the capacity limits of 4-PAM are 0.29, 0.80, 1.60 dB for
$N=100000,10000,2000$, respectively. For 8-PAM, the gaps are 0.33, 0.84,
1.60 dB, respectively. The performance of D-IRA ring codes with different
lengths shows that the optimized parameters under long-length codes also
perform well on medium-length codes.

\subsection{Examples of D-IRA Ring codes for Multi-user Networks}

\subsubsection{D-IRA ring coded CF}

\begin{figure}[h]
\centering
\includegraphics[scale=0.20]{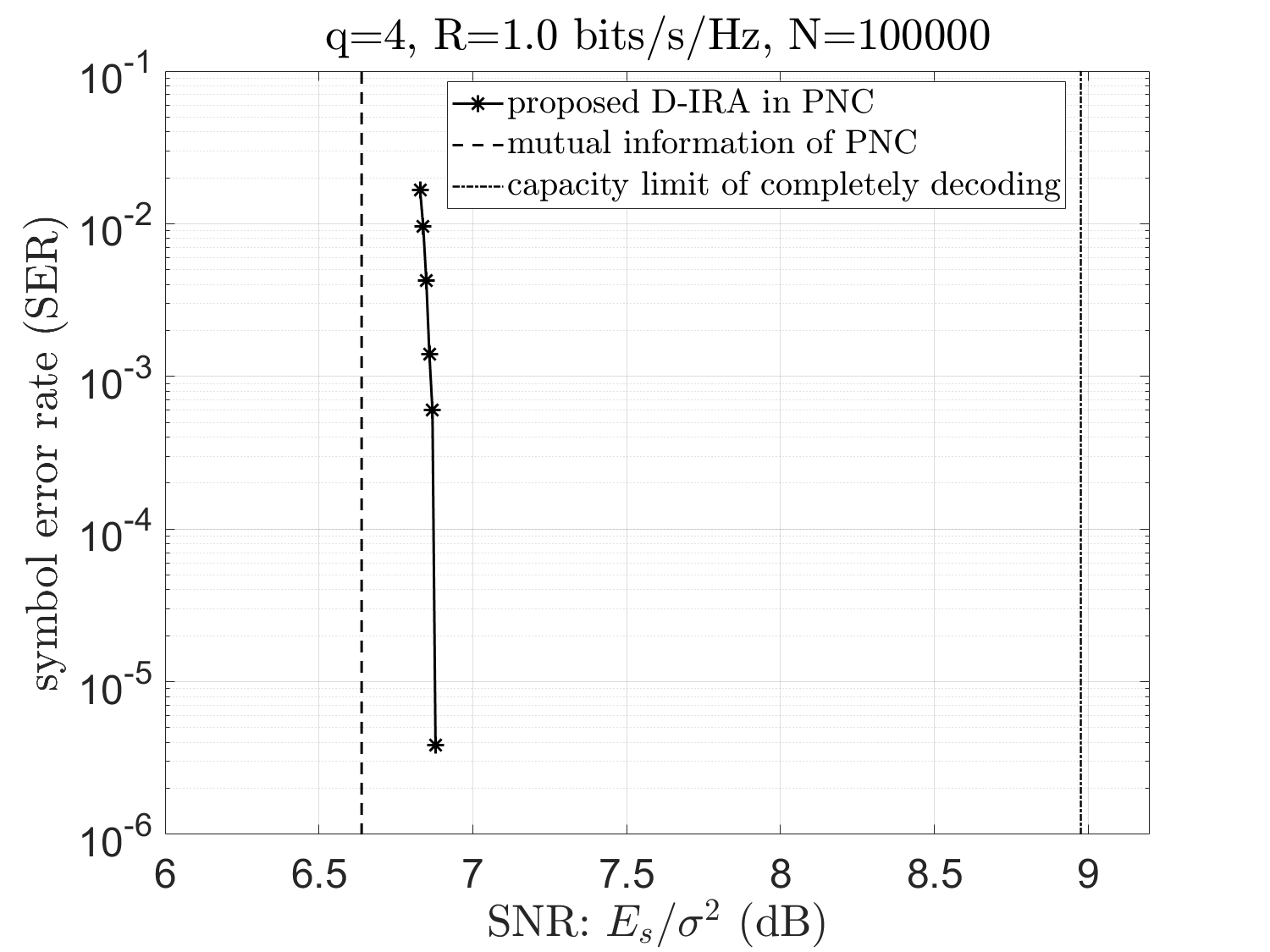}
\caption{Error-rate performance of the proposed D-IRA ring coded two-user CF
with 4-PAM modulation.}
\label{fig10}
\end{figure}

Fig. \ref{fig10} shows the error-rate performance of the proposed D-IRA ring
coded CF scheme with two users, with $4$-PAM signaling. For clarity of
presentation, we first consider the simplest example of $h_{1}=h_{2}=1$ and $%
\alpha _{1}=\alpha _{2}=1$. It can be seen that at SER of $10^{-5}$, the
performance achieved by the proposed scheme is only 0.24 dB away from the CF
mutual information \cite{nazer2007computation,lim2018joint,8066336}. The
dashed line on the right side is the capacity limit w.r.t. completely
decoding, obtained from the capacity region of two-user multiple-access
channel\cite{tse2005fundamentals}. The proposed D-IRA ring coded CF has an
advantage of at least 2.34 dB. The performance advantage becomes greater for
higher level of $2^{m}$-PAM.

\begin{figure}[h]
\centering
\includegraphics[scale=0.14]{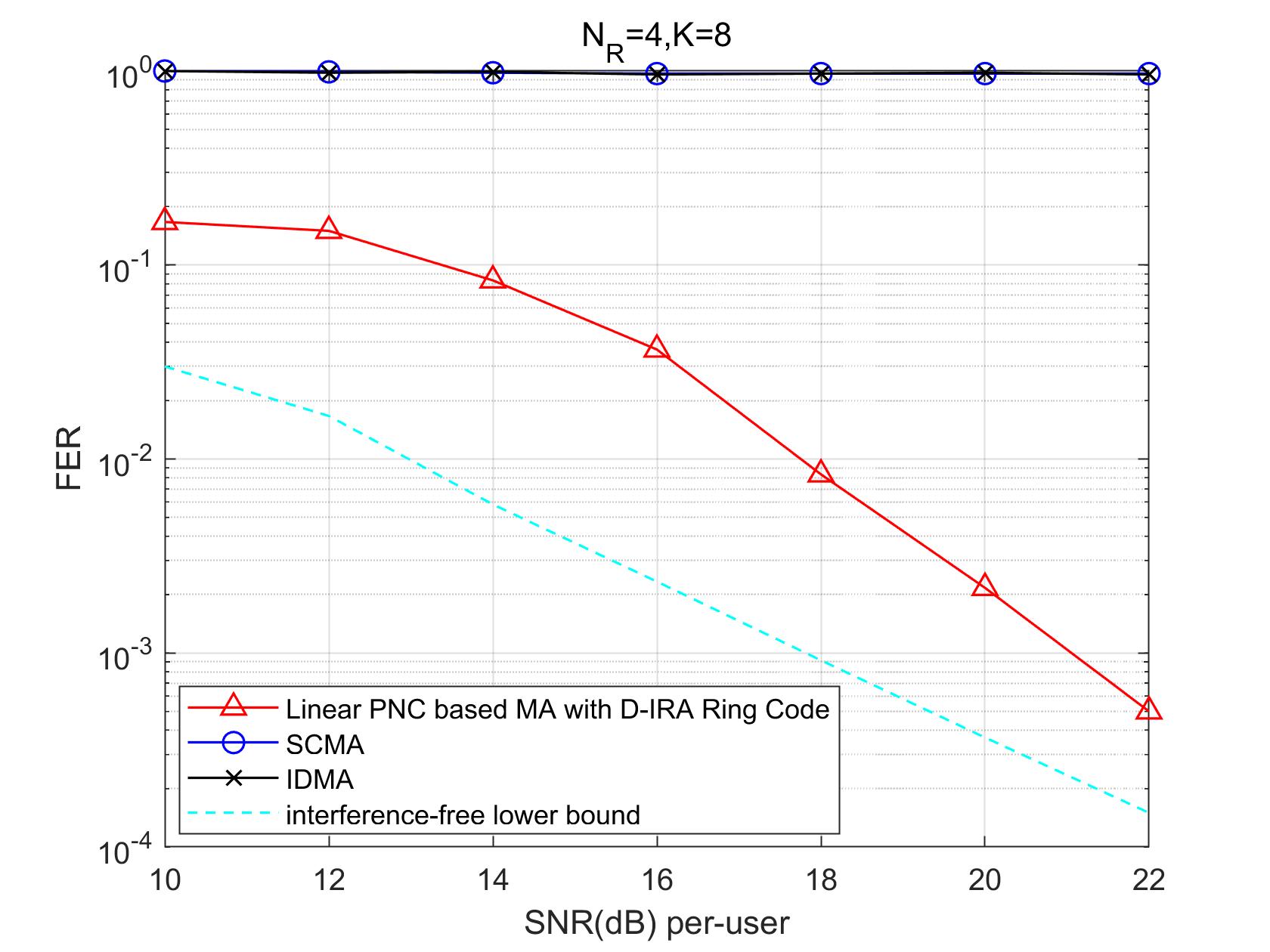}
\caption{FER of PNC (CF) based MA with proposed D-IRA
ring code of 4-PAM signaling. The FERs of IDMA
and SCMA do not decrease as SNR increases, as their iterative receivers fail
to address the interference in this overloaded case. In contrast, the CF
based MA scheme with D-IRA code performs within 3 dB the capacity limit.
Further improvement can be achieved by introducing successive computation. }
\label{fig_LCMA}
\end{figure}

We further consider a system with a larger $K$ over a Rayleigh fading
channel. Fig. \ref{fig_LCMA} shows the frame error rate (FER) of a linear
PNC (CF) based MA scheme with the proposed D-IRA ring code of 4-PAM
signaling. The number of users are $K=8$ and there are $N_R=4$ receive
antennas, where the system load is 200\%. The receiver is set compute $L=K=8$
linear message combinations in parallel, where the symbol-wise APP
calculation is extended to the multi-antenna. It is demonstrated that with
D-IRA ring codes, structuring binning can be exploited in this overloaded MA
setup. This leads to remarkable improvement over existing
interleave-division MA (IDMA) an sparse-code MA (SCMA). Further improvement
can be achieved by introducing successive computation, whose details can be
found in \cite{YangTWC22}.

\subsubsection{D-IRA ring coded DPC}


%
\begin{figure}[h]
\centering\includegraphics[width=3.5in]{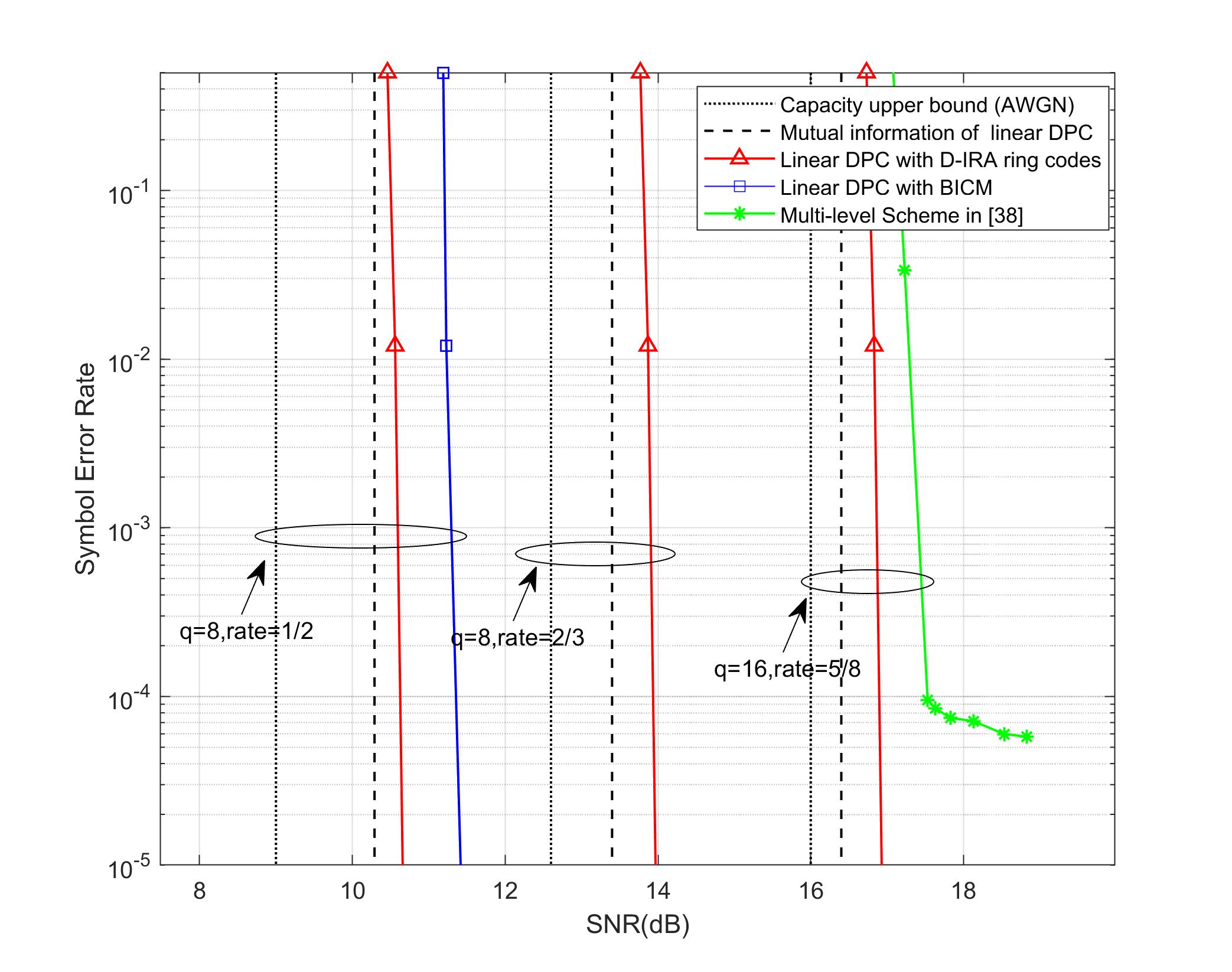}
\caption{Error rate performance of the proposed linear DPC with D-IRA ring
codes.}
\label{FigBERq8}
\end{figure}
Fig. \ref{FigBERq8} plots the error rate performance of D-IRA ring coded DPC
with $n$=50000. Our developed scheme exhibits gaps to the interference-free
AWGN channel capacity upper bound by 1.36 and 0.91dB for rates 2 and 2.5
bits/symbol, respectively. We also include the performance of the BICM based
DPC and multi-level based DPC scheme \cite{UppalTcom13}, where 4 levels are
used with average coding rate of 5/8 per-level. The proposed scheme exhibits
a 0.7 dB advantage at error rate of $10^{-4}$. Similar observations are also
observed for medium-length codes. Note that the implementation of the
presented DPC is much simpler than the multi-level design in \cite%
{UppalTcom13}.

\subsection{Complexity}


Albeit the purpose of this work is not for complexity reduction, here we
evaluate the complexity of the proposed scheme for comparison purpose. We
consider the decoding complexity of the D-IRA ring codes with FFT
acceleration, whose details are given in Appendix. The belief propagation
procedure in the iterative decoding constitutes the majority of complexity
of the D-IRA codes, hence the complexity of the calculation of symbol-wise
channel-intrinsic APPs in (\ref{Eq_symbolwiseAPP}) is not evaluated.
Following the convention, we only consider the complexity of the
multiplication operations. Denote the length of the interleaver by $\Upsilon
$. There are totally $\Upsilon +2n$ edges w.r.t. the CNs. For each edge,
there are $2\cdot 2^{m}\log _{2}{(2}^{m}{)}$ and $2^{m}\log _{2}{({2}^{m})}$
multiplications in FFT and IFFT separately. The calculated messages from FFT
are multiplied, which require $4E[d_{c,i}^{2}]n{2}^{m}$ multiplications in
total. The VNs require $E[d_{c,i}^{2}]n{2}^{m}$ multiplications.

Next consider BICM-ID with $2^{m}$ PAM. The number of outer-loop iterations
for BICM-ID receiver is denoted by $\Omega $. For fair comparison, consider
that the degree distribution of the binary IRA code is identical to that of
the $2^{m}$ D-IRA code. BICM requires $m\Omega $ binary
decoding operations, where the order of complexity of each decoding is $%
O\left( 2m\right) $. In contrast, a D-IRA ring code needs one $2^{m}$-ary
decoding, where the order of complexity is $O\left(
2^{m}\right) $. For various $m$ and $\Omega $ values, the complexity ratios
between them are shown in Table. \ref{table_8}. The
complexity of D-IRA ring codes is smaller than that of BICM with more than 4
outer-loop receiver iterations (up to 32-PAM). As the modulation level $%
2^{m} $ increases, the number of outer-loop iterations required to approach
the near-capacity also increases in BICM-ID. Comparing to the GF$\left(
2^{m}\right) $ IRA modulation codes, the D-IRA ring codes have exactly the
same order of complexity. {\small
\begin{table}[h]
\caption{The ratios between the complexities of D-IRA ring codes and BICM.}
\label{table_8}\center
{\small
\begin{tabular}{|c|c|c|c|c|}
\hline
& 4-PAM & 8-PAM & 16-PAM & 32-PAM \\ \hline
$\Omega$=4 & $25.0\%$ & $33.3\%$ & $50.0\%$ & $80.0\%$ \\ \hline
$\Omega$=6 & $16.7\%$ & $22.2\%$ & $33.3\%$ & $53.3\%$ \\ \hline
$\Omega$=8 & $12.5\%$ & $16.7\%$ & $25.0\%$ & $40.0\%$ \\ \hline
\end{tabular}
}
\end{table}
}

\subsection{Discussion of Ring Code Design for Medium and Short Packet Length%
}

Previously, we showed that code profiles optimized for long codes are also
competitive for medium-length-codes. A crucial subsequent future work along
this research direction would be to investigate how to exploit the
doubly-irregular structure in designing ring codes for scenarios with short
packet length, e.g., the messages length $k$ being 64, 128, $\cdots $, 512.
The design for such scenario will be based on optimization of the Euclidean
distance spectrum of the $2^{m}$-ary ring code with $2^{m}$-PAM signaling,
rather than the convergence behavior for long code presented in this paper.
One possible way to address this task would be to use an \textquotedblleft
error pattern impulse\textquotedblright\ based method \cite{6199937} to
optimize the minimum distance of the ring code with doubly-irregular
structure. Due to the nature of the $2^{m}$-ary processing, the minimum
girth is expected to be increased relative to conventional binary based
coded modulation schemes, and thus extremely low error-floor, e.g. frame
error probability less then $10^{-7}$ is expected to be achieved. This
future work is an interesting but quite challenging task, and is out of the
scope of the current paper.

\section{Conclusions}

%

This paper developed doubly irregular repeat accumulate (D-IRA) ring codes for $2^{m}$-PAM signaling. The proposed practical ring
codes feature the integer additive property of lattice codes. The irregular
multipliers and irregular node degree distribution, and partial random
interleavers, were designed to optimize the code profile. Numerical results
demonstrated near-capacity in point-to-point channel without the
need of outer-loop receiver iteration, as well as significant gains in
multi-user networks with compute-forward and dirty paper coding. The
proposed D-IRA ring codes provide a bridge between advanced notions in
network information theory and practical multi-user networks.
The development of D-IRA ring codes to realize other network information
theory notions, such as Slepian-wolf coding, index coding, integer-forcing,
etc., deserves further research efforts. The design of D-IRA ring codes of
short code length requires the optimization of Euclidean distance spectrum,
which is a challenging task to be studied in the future.


\section*{Appendix I \ \ FFT Accelerated Calculation for Check Nodes}

Let the vector ${\bm x}=(x_0,\cdots,x_{q-1})$. We define the vector of its
multiplication cycle as
\begin{equation}
{\bm x}^{\times g}\triangleq (x_0,x_g,x_{2\otimes g},\cdots,x_{(q-1)\otimes
g}).  \label{E1_1}
\end{equation}

According to \cite{Bennatan2006Design,4155120}, (\ref{E3}) can be rewritten
with $q$-dimension DFT/IDFT transform for prime $q$:%
\begin{equation}
\bar{\bm l}=\mathrm{IDFT}\left(\prod_{n=1}^{d-1}{\mathrm{DFT}(\bar{\bm r}^n)}%
\right),  \label{E5}
\end{equation}
where $\bar{\bm r}^n=\left({\bm r}^n\right)^{\times h_n^{-1}}$ and the
updated probability vector ${\bm l}=\bar{\bm l}^{\times (-h_d)}$.

For non-prime $q$, the superscript $h_n^{-1}$ does not exist if $h_n$ is a
zero divisor. Reconsider expression (\ref{E3}) and treat one term on the
left of equation $\sum_{n=1}^{d-1}h_n a_n=-h_{d}i$ as a whole, i.e. $h_s
a_s=j$. Then we have
\begin{equation}
l_i=\sum_{\QATOP{a_1,\dots,a_{s-1},a_{s+1},\dots,a_{d-1}\in\mathbb{Z}_q,}{%
\sum_{n=1,n\neq s}^{d-1}h_n a_n+j=-h_{d}i}}\left( \hat{r}_j^s\prod_{n=1,n%
\neq s}^{d-1} r_{a_n}^n\right),
\end{equation}
where $\hat{r}_j^s=\sum\limits_{a_s\in\mathbb{Z}_q,h_s\otimes
a_s=j}r_{a_s}^s $. Therefore, a CN input-edge with probability vector ${\bm r%
}^s$ and multiplier $h_s$ is equivalent to having vector $\hat{\bm r}^s$ and
multiplier $1$. By equivalent substitution, there is $\bar{\bm r}^n=\left(%
\hat{\bm r}^n\right)^{\times 1^{-1}}=\hat{\bm r}^n$ in expression (\ref{E5}%
), which does not require the inverse of $h_n$. The expression is applicable
to the case where some multipliers are zero-divisors, that is, the DFT
accelerated calculation is also available for the probability generation of
D-IRA ring codes. For the case of $q=2^m$, FFT and IFFT algorithms can also
be used to replace the calculation of DFT and IDFT, which further reduces
the amount of calculation.

\bibliographystyle{ieeetr}
\bibliography{RefFangtaoYu}

\end{document}